\documentstyle[prd,aps,eqsecnum,preprint]{revtex}
\input{psfig.sty}
\begin{document}
\title{Relativistic Many-Body Hamiltonian
Approach to Mesons}
\author{Felipe J. Llanes-Estrada and Stephen R. Cotanch}
\address{Department of Physics, North Carolina State
University,
        Raleigh, NC 27695-8202}
\date{\today}
\maketitle
\vspace{-2.5cm}

\begin{abstract}
   We represent QCD at the hadronic scale by means of an
effective Hamiltonian, $H$, formulated in the Coulomb gauge.  As in the
Nambu-Jona-Lasinio model, chiral symmetry is explicity broken, however our
approach is renormalizable and  also includes confinement
through a linear potential with slope specified by lattice
gauge theory.  This interaction generates
an infrared integrable singularity and we detail the computationally
intensive procedure necessary for numerical solution.  We
focus upon applications for the $u, d, s$ and $c$ quark flavors and
compute the
mass spectrum for the pseudoscalar, scalar and vector mesons.
We also perform a comparative study of alternative many-body techniques for
approximately diagonalizing
$H$:  BCS for the vacuum ground state; TDA and
RPA for the excited hadron states. The Dirac structure
of the field theoretical Hamiltonian naturally generates spin-dependent
interactions, including tensor, spin-orbit and hyperfine, and we clarify
the degree of level splitting due to both spin and chiral symmetry effects.
Significantly, we find that roughly two-thirds of the $\pi$-$\rho$ mass
difference is due to chiral symmetry and that only the RPA preserves  chiral
symmetry.  We also document how hadronic mass  scales are generated by
chiral symmetry breaking in the model vacuum.  In  addition to the vacuum
condensates, we compute meson decay constants and detail the
Nambu-Goldstone realization of chiral symmetry by numerically verifying 
the  Gell-Mann-Oaks-Renner relation. Finally, by including D
waves in our charmonium calculation we have resolved the anomalous
overpopulation 
of $J/\Psi$ states relative to observation.
\\*[.2 cm]
\hspace{-1.5em}PAC number(s): 12.39.Pn, 12.40.Yx
\end{abstract}

\pagebreak
\section{Introduction}

In a series of publications \cite{ssjc96,robertson,flesrc,gjc} an ambitious QCD
program has been initiated to comprehensively investigate
hadron structure.  The theoretical formulation entails
renormalization and utilizes established
many-body techniques to approximately diagonalize an effective confining
Hamiltonian.
This paper, a detailed continuation of our recent letter \cite{flesrc},
focuses upon the quark sector
and reports numerical results for mesons complementing our previous
gluon study \cite{ssjc96}.

Over the years there have been many meson
investigations, from the early, simple
non-relativistic constituent quark model calculations to more involved
relativistic, field theoretical approaches implementing current quarks
and spontaneous chiral symmetry breaking. A
common shortcoming of these analyses is an inability to consistently
reproduce the physical mass spectrum of the scalar and pseudoscalar mesons.
Our paper addresses this issue and significantly extends the pioneering
work of the Orsay group \cite{Orsay}, Adler and Davis \cite{Adler}, and
the Lisbon investigators \cite{Lisbon}.  In our approach the exact QCD
Hamiltonian in the Coulomb gauge is modeled by an effective,
confining Hamiltonian, $H$,  that is fully
relativistic with quark field operators and current quark
masses.  However, before approximately diagonalizing $H$, a similarity
transformation is implemented to a new quasiparticle basis having a
dressed, but unknown constituent mass.  As described in Sec. II, this
transformation entails a rotation which mixes the bare quark creation and
annihilation operators.  By then performing a variational calculation
to minimize the ground state (vacuum) energy, a specific angle and
corresponding quasiparticle mass is selected.  In this fashion chiral
symmetry is dynamically broken and a non-trivial vacuum with quark
condensates emerges.  This treatment is precisely analogous to
the Bardeen, Cooper, and Schrieffer (BCS) description of a superconducting
metal as a
coherent vacuum state of interacting quasiparticles combining to form
condensates
(Cooper pairs). Excited states (mesons) can then be represented as
quasiparticle
excitations using standard many-body techniques which in this work will
be the
Tamm-Dancoff (TDA) and random phase approximation (RPA) methods.
The two treatments are truncated at the one
quasiparticle, one quasihole level and then numerically compared.
Our RPA analysis confirms and extends the early work of Ref. \cite{weise}
which utilized an extended Nambu-Jona-Lasinio mean field approach.

Two other comments are in order before proceeding.  First, there are
several reasons for choosing the Coulomb gauge framework.  As discussed by
Zwanziger
\cite{zwanziger}, the Hamiltonian is renormalizable in this gauge and,
equally as important, the Gribov problem (${\bf\nabla}\cdot{\bf A}=0$
does not uniquely specify the gauge) can be resolved (see Refs.
\cite{robertson,zwanziger} for further discussion).   Related,
there are no spurious gluon degrees of freedom since only transverse
gluons enter.  This ensures all Hilbert vectors have positive normalizations
which is essential for using
variational techniques that have been widely successful
in atomic, molecular and  condensed
matter physics.
Second, due to Fock space truncations our analysis is not
Lorentz invariant. However, we only plan
to use one frame and do not
compute hadron form factors with this method. Interestingly,
violating Lorentz non-invariance implies a prefered reference frame,
which, as selected by chiral symmetry breaking, is the condensate rest
frame.

This paper consists of six sections and two appendices.  The next section
introduces our effective, QCD inspired Hamiltionian and developes the BCS
vacuum treatment leading to the quasiparticle mass gap equation.
We also compare our approach to the classic
Nambu-Jona-Lasinio model.  In Sec. III we detail our
numerical, supercomputer solution of the gap equation along with the
quark condensate and constituent mass values.  Sections IV A and IV B
describe the TDA and RPA, respectively, while Sec. IV C addresses weak decays
and Sec IV D presents a derivation of the Gell-Mann-Oakes-Renner relation.
The TDA and RPA
meson spectra are compared and discussed in Sec. V.  This section
also includes results from a simple $SU_f(3)$ flavor mixing analysis for the
$\eta$-$\eta'$ system and our predictions for the charmed mesons.  Conclusions and
future work are summarized in Sec. VI.  Finally, Appendix A provides further details
regarding  the BCS transformation and vacuum state while Appendix B presents the
most general TDA equation for arbitrary angular momentum.

\section {Hamiltonian and mass gap equation}

\subsection{Effective Hamiltonian}

By introducing a phenomenological confining potential, $V_L$, the QCD Coulomb
gauge Hamiltonian \cite{robertson} for the quark sector can be replaced by
an effective Hamiltonian

\begin{equation}
\label{Hamiltonian} H = \int d \vec{x} \Psi ^{\dagger}
(\vec{x}) (-i\vec{\alpha}\cdot \vec{\nabla} + \beta m )
\Psi(\vec{x}) - \frac{1}{2} \int d \vec{x} d \vec{y}
\rho^a(\vec{x})V_L(\arrowvert \vec{x} -
\vec{y} \arrowvert) \rho^a(\vec{y})
\end{equation}
where $\Psi$, $m$ and $\rho^a(\vec{x}) =
\Psi^{\dagger}(\vec{x})T^a\Psi(\vec{x})$
are the current (bare) quark field, mass and color density,
respectively (for a more complete discussion, especially for the heavy
quark sector, consult Refs.  \cite{zwanziger,Nora}).  For notational ease
the flavor subscript is omitted (same $H$ for each flavor) and the color
index
runs $a = 1...8$.
Motivated by lattice gauge studies we adopt a linear confining interaction,
$V_L = \sigma\arrowvert\vec{x} -\vec{y}\arrowvert$, with slope
$\sigma = .18$ $GeV^2$ also specified by lattice and Regge phenomenology.
In our
analysis we have also performed calculations with and without the leading QCD
canonical or
Coulomb (one-gluon exchange) interaction, $V_C =
-\frac{\alpha_s}{\arrowvert\vec{x}
-\vec{y}\arrowvert}$, with $\alpha_s = \frac{g^2}{4\pi}$ $\cong$ $.4$.   For
most observables, especially the meson mass spectrum, the Coulomb
interaction is
not important and can be omitted.  This can be understood by noting that in
momentum space, where we perform all calculations, the two interactions
have the
same sign, i.e. $V(r=|\vec{x}-\vec{y}|) = V_C + V_L$

\begin{eqnarray}
\hat{V} (k) &=& \int
d\vec{r}\ V(r)e^{-i\vec{k}\cdot\vec{r}}\nonumber\\
 &=& -4\pi \frac{\alpha_s}{k^2} - 8\pi \frac{\sigma}{k^4} \ .
\end{eqnarray}
Because the meson wavefunctions have a finite momentum distribution,
most static meson properties are predominantly governed by the
infrared (IR), or low, momentum region where the confining potential
dominates.
Including the Coulomb interaction is then roughly equivalent to using a
slightly larger string tension, $\sigma$.  There are certain observables,
and in
particular the gap equation detailed below, for which the Coulomb interaction
is ultra-violet (UV) divergent.  In such cases we regularize with a
cut-off parameter and then could renormalize to remove cut-off sensitivity
using one of our  renormalization procedures detailed in Refs.
\cite{robertson,gjc} for the gluon sector.  In this paper we only present
unrenormalized results since this program is still in progress
\cite{NCSU4,gjc2}
and has not yet been completed for the quark sector.  This is an additional
reason for omitting the Coulomb interaction.  Hence, with the exception of the
current quark masses  (we use $m_u = m_d = 5 $ $MeV$, $m_s = 150$ $MeV$,
$m_c = 1200$ $MeV$), our
approach entails only one pre-determined parameter which also sets the
hadronic scale, $\sqrt{\sigma} = 424$ $MeV$.

We further note that even though the confining potential is IR divergent,
this singularity is cancelled (see Ref.
\cite{Adler}) in both the mass gap equation and all calculations for associated
observables. Hence, the problem is the delicate numerical evaluation of this
integrable singularity which we discuss in Sec. III.

Finally, we stress that in constituent quark models free
quarks can exist which requires imposing color confinement.
However, as demonstrated in Refs. \cite{Orsay,Adler,Lisbon} the
Lorentz structure of our Coulomb
gauge density-density confining
interaction only permits stable solutions for color singlet states.
Therefore, confinement naturally emerges in our approach.

\subsection{BCS transformation and gap equation}

We now wish to solve $H \Psi = E \Psi$ as accurately as possible.
In this subsection we focus on the ground state and introduce the
Bogoliubov-Valatin, or BCS, transformation.  We begin by recalling the
plane wave, spinor expansion for the quark field operator

\begin{equation} \Psi(\vec{x})=\sum_{c \lambda}
\int \frac{d\vec{k}}{(2\pi)^3}
\left[u_{c\lambda}(\vec{k})b_{c\lambda}(\vec{k})
+ v_{c\lambda}(-\vec{k})d^{\dagger}_{c\lambda}(-\vec{k}) \right] e^{i\vec{k}
\cdot \vec{x}} \end{equation}
with free particle, anti-particle spinors $u_{c\lambda}, v_{c\lambda}$ and
bare creation,
annihilation operators $b_{c\lambda}, d_{c\lambda}$ for current quarks,
respectively.
Here the spin state (helicity) is denoted by $\lambda$ and color index by
$c = 1,2,3$
(which is hereafter suppressed). Because we can expand $\Psi$ in terms of any
complete basis we may equally well use a new quasiparticle basis
\begin{equation} \Psi(\vec{x})=\sum_{\lambda}
\int \frac{d\vec{k}}{(2\pi)^3}
\left[U_{\lambda}(\vec{k})B_{\lambda}(\vec{k})
+ V_{\lambda}(-\vec{k})D^{\dagger}_{\lambda}(-\vec{k}) \right] e^{i\vec{k}
\cdot \vec{x}} \end{equation}
entailing quasiparticle spinors $U_{\lambda}, V_{\lambda}$ and operators
$B_{\lambda}, D_{\lambda}$.
The Hamiltonian is equivalent in either basis and the two are related
by a similarity (Bogoliubov-Valatin or BCS) transformation.  The
transformation
between operators is given by the rotation
\begin{equation}\label{BVrotation}
 B_{ \lambda }(\vec{k}\ ) = \cos \frac{\theta_k}{2} b_{
\lambda }(\vec{k}\ )
-\lambda \sin \frac{\theta_k}{2} d^{\dagger}_{ \lambda}(-\vec{k}\ )
\end{equation}
$$D_{ \lambda }(-\vec{k}\ ) = \cos \frac{\theta_k}{2} d_{
\lambda}(-\vec{k}\ )
 +\lambda \sin \frac{\theta_k}{2} b^{\dagger}_{ \lambda }(\vec{k}\ )$$
involving the BCS angle $\theta_k = \theta(k)$.  Similarly the rotated
quasiparticle
spinors are
\begin{equation} \label{inverseBV}
U_{\lambda}(\vec{k}) = \cos \frac{\theta_k}{2}
u_{\lambda}(\vec{k})
- \lambda \sin \frac{\theta_k}{2} v_{\lambda}(-\vec{k})= \frac{1}
{\sqrt{2}}
\left[ \begin{array}{c} \sqrt{1+\sin\phi(k)}\:\:
\chi_{\lambda}    \\ \sqrt{1-\sin \phi(k)}\:\:
\vec{\sigma} \cdot \hat{k}\:\: \chi_{\lambda}
\end{array} \right]
\end{equation}
$$\quad V_{\lambda}(-\vec{k}) = \cos \frac{\theta_k}{2} v_{\lambda}(-\vec{k})
+ \lambda \sin \frac{\theta_k}{2} u_{\lambda}(\vec{k})
 = \frac{1}{\sqrt{2}}\left[
\begin{array}{c} -\sqrt{1-\sin \phi(k)}
\:\:\vec{\sigma}\cdot \hat{k} \:\:\chi_{\lambda} \\
\sqrt{1+\sin \phi(k)} \:\:\chi_{\lambda}\end{array}
\right]$$
where $\chi_{\lambda}$ is the standard two-dimensional Pauli spinor.  We have
also introduced the gap angle,
$\phi_k =\phi (k)$, which is related to the BCS angle, $\theta /2$,
by $\phi = \theta + \alpha$ where $\alpha$ is the current, or perturbative,
mass angle satisfying $\sin  \alpha = m/E_k$ with $E_k = \sqrt {m^2 + k^2}$.
Hence
$$\sin\phi_k = \frac{m}{E_k} \cos \theta_k + \frac{k}{E_k} \sin \theta_k$$
$$\cos\phi_k= \frac{k}{E_k} \cos \theta_k - \frac{m}{E_k} \sin \theta_k \ .$$

Similarly, the perturbative, trivial vacuum, defined by
$b_{ \lambda } \arrowvert 0 \rangle = d_{ \lambda } \arrowvert 0 \rangle  =
0$, is related to the quasiparticle vacuum, $B_{ \lambda } \arrowvert \Omega
\rangle = D_{
\lambda } \arrowvert
\Omega \rangle  = 0$,  by the transformation
\begin{equation} \label{BCS} \arrowvert \Omega \rangle = \exp \left(
-\sum_{\lambda } \int \frac{d\vec{k}}{(2\pi)^3}
\lambda \tan\frac{\theta_k}{2}b^{\dagger}_{ \lambda }(\vec{k})
d^{\dagger}_{ \lambda }(-\vec{k})  \right) \arrowvert 0 \rangle \ .
\end{equation}
In this paper we will denote the BCS vacuum by $\arrowvert \Omega
\rangle$ (in Sec. IV we introduce the RPA vacuum labeled
$\arrowvert \Omega_{RPA}\rangle $).  Expanding the exponential and noting that
the form of the operator
$b^{\dagger} d^{\dagger}$  is designed to create a current
quark/antiquark pair with the vacuum quantum numbers, clearly exhibits the
BCS vacuum as a coherent state of quark/antiquark excitations
(Cooper pairs)
representing $  ^{2S+1}L_J  = $${^{3}}$P$_0$  condensates.
One can regard
$\tan \frac{\theta_k}{2}$ as the momentum wavefunction of the pair in
the  center of momentum system.

We now seek an approximate ground state for our effective Hamiltonian by
minimizing the BCS vacuum expectation, $\langle \Omega \arrowvert H
\arrowvert \Omega \rangle $.  We do this variationally using the gap angle,
$\phi_k$, (not the BCS angle) which leads to the gap equation,
$\delta \langle \Omega \arrowvert H \arrowvert \Omega \rangle =0$.
After considerable mathematical reduction, the nonlinear integral
gap equation follows
\begin{equation} \label{gap3d}
k\sin\phi_k -m\cos\phi_k = \frac{2}{3} \int \frac{d\vec{q}}{(2\pi)^3}
\hat{V}(\arrowvert \vec{k}
-\vec{q} \arrowvert ) [\sin \phi_k \cos \phi_q \hat{k} \cdot \hat{q} -
\sin \phi_q \cos \phi_k] \ . \end{equation}
The angular integrals can be analytically evaluated (see Appendix B) to give

\begin{equation} \label{gap1d}
 k\sin \phi_k -m \cos \phi_k =\frac{2}{3} \frac{1}{(2\pi)^2}
\int_0^{\infty} q^2dq [\cos \phi_q \sin \phi_k \hat{V}_1(k,q) - \cos \phi_k
\sin \phi_q \hat{V}_0(k,q)]
\end{equation}
where $$ \hat{V}_0=\frac{-16\pi \sigma}{(k^2-q^2)^2}$$
$$\hat{V}_1=\frac{2\pi\sigma}{k^2 q^2} \left[ ln \left( \frac{k+q}{k-q}
\right) ^2 +
(k^2+q^2)\left( \frac{-4qk}{(k^2-q^2)^2} \right) \right] $$
corresponding to the linear potential above.  Similar expressions for the
Coulomb potential are given in Appendix B.

There are several alternative ways to derive this same gap equation.  One is
through the Ward identites. Another is by requiring
cancellation of the anomalous
Bogoliubov terms in the 2-body part of the newly normal ordered
Hamiltonian.  The latter is necessary to stabilize the vacuum and is also
equivalent to minimizing the 0-body
constant energy splitting the BCS and trivial vacua
(see Ref. \cite{Lisbon}).

Numerically we actually  solve a different form of the gap equation,
originally obtained by
Adler and Davis \cite {Adler},
that is more familiar to the solid state
community.
They use the function $\psi_k = \psi (k)$ related to our gap angle by
$$\sin \phi_k = \frac{2 \psi_k}{1+\psi^2_k}$$
$$\cos  \phi_k = \frac{1-\psi^2_k}{1+\psi^2_k}$$
with corresponding gap equation
\begin{equation} \label{GAP1D}
 k\psi_k -\frac{m}{2}(1-\psi^2_k)= \frac{2}{3(2\pi)^2}
\int_0^{\infty} q^2 dq
\frac{ \hat{V}_1 \psi_k(1-\psi^2_q) - \hat{V}_0 \psi_q
(1-\psi^2_k)}{1+\psi^2_q} \ .
\end{equation}

Examination of Eqs. (\ref{gap1d},\ref{GAP1D}) reveals that
the divergence at $k = q$ is an integrable singularity
for the linear potential since for all $k$ 
the integrands vanish at $k = q$.
This is not the case for the  Coulomb potential UV singularity. However,
since it naturally emerges from the canonical QCD Hamiltonian (one gluon
exchange), we  retain the option of including this
 potential
for selected calculations and
use a cut-off to regulate its
ultraviolet (UV) divergence.

The solution of the gap equation (see Sec. III)
leads to a vacuum quark-antiquark condensate given by

\begin{equation} \label{cond}
\langle \overline{q}q \rangle\equiv\langle \Omega \arrowvert \overline{\Psi}(0)
\Psi(0)
\arrowvert \Omega \rangle
= - \frac{3} {\pi^2} \int  k^2 \sin \phi_k  \  d{k} \end{equation}
which is quadratically divergent for non-zero current quark mass $m
\neq 0$.
We regulate this by subtracting the trivial
condensate contribution giving

\begin{equation} \label{cond1}
\langle \overline{q}q \rangle_{reg}  = - \frac{3} {\pi^2} \int
k^2 \left(  \sin \phi_k
-\frac{m}{E_k}  \right)\ dk \ .
\end{equation}

Our model is color confining and does not permit free solitary particles
since the self-energy or dispersion relation

\begin{equation}\label{dispersion}
\epsilon_k = m\sin\phi_k + k\cos
\phi_k - \frac{2}{3}
\int \frac{d\vec{q}}{(2\pi)^3} \hat{V}(\arrowvert
\vec{k}-\vec{q} \arrowvert)(\sin\phi_k
\sin \phi_q + \hat{k} \cdot \hat{q} \cos \phi_k \cos \phi_q )
\end{equation}
obtained
from the 1-body part of $H$, Eq. (\ref{Hamiltonian}),
is divergent (now there is no cancellation at the
singular point $\vec{k} = \vec{q}$).
Further, this divergence is also cancelled in the
bound state equation but only for color singlet states (see below).
Even though the self-energy diverges it is still useful to introduce
the concept of an effective quasiparticle (constituent) mass,  ${\sc m}_q $,
which can be extracted from the low momentum behavior of
the gap angle.  We introduce a running, dynamical mass, ${\sc m}(k)$, by an
effective Dirac spinor in canonical form
\begin{equation}
U_{\lambda}^{eff}(\vec{k}) = N
\left[ \begin{array}{c} \:\:
\chi_{\lambda}    \\ \:\:
\frac{\vec{\sigma} \cdot \vec {k}}{E + {\sc m}(k)}\:\: \chi_{\lambda}
\end{array} \right]
\end{equation}
with normalization $N$ and $E = \sqrt{{\sc m}^2(k) + k^2}$.  Then using this
equation and Eq.
(\ref{inverseBV}) we equate the two relative normalizations between upper and
lower spinor components yielding a relation between the running dynamical mass
and  gap angle
\begin{equation}
\frac{\sqrt{1+\sin\phi(k)}} {\sqrt{1-\sin\phi(k)}} = \frac{E + {\sc m}(k)}
{k},
\end{equation}
or
\begin{equation}
\sin\phi(k) = \frac {{\sc m}(k)} {E} = 1 - \frac {k^2} { 2 {\sc m}^2(k)}   +
\vartheta (k^4) .
\end{equation}
We identify the dressed quark or quasiparticle mass as ${\sc m}_q =  maxima \, 
{\sc m}(k)$ and extract it
from the low momentum behavior of the gap angle (essentially inverse of the
slope
near zero momentum).
The value of ${\sc m}_q$ characterizes the degree of chiral symmetry breaking
and can be loosely regarded as the constituent quark mass
associated with phenomenological quark models.

Note that our expression for the running mass is functionally
identical to the perturbative expression
$\tan\alpha =
\frac{m}{k}$.  Related, since the rotated quasiparticle spinors have a running
momentum dependence, they no longer rigorously
provide a representation of the Lorentz group.  Our form of dynamical chiral
symmetry breaking violates Lorentz invariance which implies a preferred
reference
frame, namely the condensate rest frame. For most static observables such
as masses, condensates and decay constants, Lorentz symmetry is not important.
However, for some observables, such as electromagnetic form factors, care is
necessary and boost corrections may be important.  This issue is
under investigation and will be reported in a future communication.

\subsection{Comparison to the Nambu-Jona-Lasinio model}

The classical effective model of Nambu-Jona-Lasinio (NJL) (see Ref.
\cite{Miransky} for review)
entails various Lagrangian formulations, a common one being
\begin{equation}
{\mathcal L}= i \bar{\Psi} \not{\!\partial} \Psi +
G[(\bar{\Psi}\Psi)( \bar{\Psi} \Psi) - (\bar{\Psi}\gamma_5 \Psi)
(\bar{\Psi} \gamma_5 \Psi)]
\end{equation}
where $G$ is a constant.
It is customary to introduce the approximations  $\bar{\Psi} \gamma_5
\Psi \approx 0$ and $\bar{\Psi} \Psi \approx \langle \bar{\Psi} \Psi
\rangle_{vacuum} $ to linearize the equations of motion and then extract
a constituent quark mass from the NJL mass gap
equation. In this fashion chiral symmetry breaking is achieved.

Our formulation extends beyond the NJL model in several important ways.
\begin {enumerate}
\item  Our approach is more general and permits explicit gluonic
degrees of  freedom (see Refs. \cite{ssjc96,robertson,gjc}). The unification
of the quark and glue sectors is crucial for a comprehensive treatment
of hadron structure, especially for glueball and hybrid meson systems.
\item  Our formulation includes confinement and is renormalizable while
the NJL model has neither.  The NJL pointlike interaction would be
recovered in the limit
$V(\vec{x}-\vec{y}) \longrightarrow
G \delta(\vec{x} - \vec{y}) $ which removes
all important nonlocalities.

\item  Our model has a  density-density interaction kernel with a
different Lorentz
structure, $\gamma_0 \gamma_0$, which is the product of
four-vector time components. As discussed in  Ref. \cite{NCSU4}, a
density-density (vector-vector) interaction is superior to the
scalar-pseudoscalar  displayed by the NJL model.

\item  The chiral symmetry breaking mode of the NJL is extremely
restrictive yielding a constant quasiparticle mass, $m_{dyn}$, and
simple dispersion $E = \sqrt{m^2_{dyn} + k^2}$.
Related, the NJL limit of our model
also yields a more restricted gap angle since
$\sin \phi = \frac {m_{dyn}} {\sqrt{m^2_{dyn}+ k^2}}$.
Our method has a running
mass and different quasiparticle dispersion which yields more realistic
TDA and RPA hadron masses.


\end{enumerate}

\section{Numerical solution of the gap equation}

The gap equation (\ref{gap1d}) has been previously solved for the harmonic
oscillator potential, where it takes its simplest form as a differential
equation (see Ref. \cite{Orsay,Lisbon}), and also for the linear potential
(see Ref. \cite{Adler,Lisbon}). Here we summarize our analysis
which confirms and extends
the latter results.

To numerically treat the integrable IR singularity a
regularization must still be implemented even though the final results
are independent of this procedure.
We considered several different regularizations.  We first tried an
analytical regulator (equivalent to a deconfining correction to the potential).
This was an unstable algorithm and and convergence could not be achieved.
We next examined the method of Ref.
\cite{Adler} which off-sets the q-discretization by
half a step in the kernel with respect to the k-discretization.
This procedure was also rejected as it was less amenable
for documenting the regulator sensitivity.
We finally adopted the simplest method of omitting the singular point
${k}={q}$.  This also facilitated
a controlled senstivity study by
just increasing the number of mesh points.
Related, we adopted a variable mesh size to integrate more efficiently
and mapped the integration variable $q$ to $v$
$$ q=\frac{v^2 q_{max}}{1+q_{max}(1-v)} $$
for N points uniformly distributed in the interval $v \in (0,1)$.

Following Ref. \cite{Adler} we  elected to solve the gap equation
in form specified by Eq. (\ref{GAP1D}) and also utilized the Gauss
algorithm as described there.  The Gauss method assures convergence
but is rather inefficient for extensive sensivity studies in parameter
space.  We therefore modified our numerical approach by first finding a good
approximate solution, $\psi^0_k$, to the non-linear gap equation and
then obtained
a linear equation for the desired correction, $\delta_k$,
giving the final solution
\begin{equation} \label{guess}
\psi_k=\psi^0_k+\delta_k
\end{equation}
to arbitrary accuracy.
Substituting Eq. (\ref{guess}) in the gap equation, Eq. (\ref{GAP1D}), and
dropping higher powers of $\delta_k$ yields the approximate linear equation

\begin{eqnarray*}
&\delta_k\left[ (k+m)\psi^0_k - \frac{2}{3(2\pi)^2} \int_0^{\infty}
\frac{q^2dq}{1+\psi^{0^2}_q} \left[ \hat{V}_1(1-\psi^{0^2}_q)+2\hat{V}_0
\psi^0_q\psi^0_k \right] \right] \\
& + \frac{2}{3(2\pi)^2} \int_0^{\infty}
\frac{q^2dq}{1+\psi_q^{0^2}} \left[ \hat{V}_0 \left( 1-\frac{2\psi_q^{0^2}}
{1+\psi^{0^2}_q} \right) (1-\psi_k^{0^2}) + 4\hat{V}_1 \frac{\psi_k^0
\psi^0_q}{1+\psi_q^{0^2}} \right]\delta_q=   \\
&= -k\psi_k^0+\frac{m}{2}(1-\psi^{0^2}_k) - \frac{2}{3(2\pi)^2}
\int_0^{\infty} \frac{q^2dq}{1+\psi_q^{0^2}} \left[ \hat{V}_0 \psi_q^0
(1-\psi_k^{0^2})-\hat{V}_1\psi_k^0(1-\psi^{0^2}_q) \right] \ .
\end{eqnarray*}
This equation is of the form \begin{equation} \int dq{\mathcal A}(k,q)
\delta(q)= {\mathcal B}(k) \end{equation}
which can be solved for $\delta_k$ by matrix inversion.
We found the Gauss algorithm sufficient for obtaining the initial approximate
solution $\psi^0_k$.  To achieve full convergence
required up to 12,000 mesh points, a factor of 60
more than the early calculations of Ref.
\cite{Adler}.

We checked our computer codes by calculating two different toy kernels
$\hat{V}_0=
\frac{k}{q^2(1+q^2)}$,  $ \hat{V}_1=0$ and $\hat{V}_0=0$,  $\hat{V}_1=
\frac{k}{q^2(1+q^2)}$, each designed to yield a known constant
value for $\psi_k$.
We then performed a series of cut-off sensitivity runs and mapped out
the convergence rate as a function of mesh point number $N$ which
ranged from 100 to 12,000.  We used the quark condensate as a test
observable and also performed calculations for zero and non-zero
current quark mass, $m$, with and without the Coulomb
potential using  $\alpha_s = .4$.  For $m = 0$, $\alpha_s = 0$, we determined the
sensitivity to $N$ (the effective cut-off parameter) was slightly higher than
previously reported
\cite{Adler,NCSU4} and given by
\begin{equation}
\langle \overline{\Psi} \Psi \rangle \simeq -\left[ \left( 113 -
\frac{1400}{N} \right) MeV \right]^3 \ .
\end{equation}
Note that this number is somewhat smaller than the commonly accepted
lattice value of about
$-(250\ MeV)^3$.
Including the Coulomb potential only increases the
condensate to $-(119\ MeV)^3$. We therefore conclude an improved model ground
state is needed which can be provided by including additional terms
in the Hamiltonian, such as the quark-gluon minimal coupling (hyperfine)
interaction. This point is also affirmed below in our RPA treatment which does
yield a more realistic condensate value.

Our other key result, which will be of interest in
connection with chiral perturbation theory \cite{Dobado,Gasser},
is for the constituent quark mass and the BCS condensate as a function of
the $u$, $d$ quark mass.  Now it is necessary to use Eq. (\ref{cond1})
and also impose an additional integration cut-off limit ($q_{max}$ around 10
$GeV$).
This yields
$$ -\langle \overline{\Psi} \Psi \rangle ^{\frac{1}{3}} = 2.03 \ m+113.1$$
$${\sc m}_q=1.6 \ m+77.9$$  where the units are $MeV$.
A precision calculation for ${\sc m}_q$ with $m=0$ yields the
slightly higher value ${\sc m}_q=$ $80.5 \ MeV$ (all values for the linear potential
only).

Finally, we note that our Hamiltonian is $SU_f(3)$ flavor symmetric,
broken only by the small current flavored quark mass term.  However,
and quite significant, the vacuum properties and gap angle exhibit
substantial $SU_f(3)$ violations as evidenced by our strange quark
calculations using a current mass of 150 $MeV$. Important violations
occur even for a strange quark mass as low as 50 $MeV$.  While this
result may be model dependent it does suggest
that certain chiral symmetry arguments in the
literature regarding the strange quark sector should be taken with care.

\section{Many-Body Techniques}

We now formulate mesons as excited states consisting of quasiparticles
and seek approximate eigensolutions of our Hamiltonian.  We first develop
the TDA and then treat the RPA
in the next subsection.  Of the two, only the RPA
preserves chiral symmetry, as we detail below.  It is therefore more
closely related to the Bethe-Salpeter formalism \cite {Argonne}
incorporating the Schwinger-Dyson quark propagator
using an instantaneous interaction
in the rainbow approximation (equivalent to our gap equation).  We will document
this connection more formally in a future publication.

\subsection{TDA equation of motion}\label{TDAsection}

The principle advantage of the TDA is that it is a controllable approximation
which truncates the Fock-space expansion for a chosen level of calculational
effort and resources.
In terms of the quasiparticle operators introduced in Sec. II,
we introduce the TDA meson creation operator
\begin{equation}\label{TDAop}
Q^{\dagger}_{nJP}(TDA) = \sum_{\gamma \delta}
\int \frac{d\vec{k}}{(2\pi)^3} \Psi ^{nJP}_{\gamma \delta}(\vec{k})
B^{\dagger}_{\gamma}(\vec{k}\ ) D^{\dagger}_{\delta} (-\vec{k}\ ) \ .
\end{equation}
A meson
with quantum numbers $nJP$ (radial-node number, $n$, total angular momentum,
$J$, and parity, $P$)
is then represented by the Fock space expansion

\begin{equation}\label{TDAansatz}
\arrowvert \Psi^{nJP}_{TDA} \rangle = Q^{\dagger}_{nJP}(TDA)\arrowvert
\Omega \rangle
\end{equation}
containing a quasiparticle and quasihole excited from the BCS vacuum.
The Hamiltonian equation
is then projected onto this 1p-1h truncated Fock sector giving
the TDA equation

\begin{equation} \label{TDA} \langle \Psi^{nJP}_{TDA} \arrowvert [{H},
B_{\alpha}^{\dagger}D_{\beta}^{\dagger}]
\arrowvert \Omega \rangle= (E_{nJP}-E_0)\Psi^{nJP}_{\alpha \beta} \ . \end{equation}
In evaluating the commutator  we note
$$ [{H}_0,B^{\dagger}_{\alpha}D^{\dagger}_{\beta}]=0$$
$$\langle \Psi^{nJP}_{TDA} \arrowvert [{H}_2
,B^{\dagger}_{\alpha}D^{\dagger}_{\beta}]\arrowvert
\Omega \rangle = (\epsilon_{k}+ \epsilon_{k})
\Psi^{nJP}_{\alpha\beta} $$
and for the two body potential
$$ \langle \Psi_{TDA}^{nJP} \arrowvert [{H}_4
,B^{\dagger}_{\alpha} D^{\dagger}_{\beta} ] \arrowvert
\Omega \rangle = \frac{4}{3} \sum_{\gamma \delta} \Psi^{{nJP}}_{\delta \gamma}
\hat{V}( {\vec k}_\delta - {\vec k}_\gamma ) U^{\dagger}_\alpha U_{\delta}   
V^{\dagger}_\gamma V_\beta 
$$
where $\epsilon_k$ is the BCS gap energy given by (\ref{dispersion})
and $H_N$ is the Hamiltonian component containing N field
operators (after normal ordering with respect to the BCS vacuum). 

We can exploit the rotational invariance of our Hamiltonian and reduce the
linear
TDA equation to a one dimensional, nonlocal equation by an angular momentum
decomposition.
Introducing the orbital and spin angular momenta
$\vec{L}$ and $\vec{S}$, respectively, the meson state vector can be
expanded in partial-waves involving a one-dimensional (radial) wavefunction
$\Psi^{nJP}_{LS}$
\begin{equation} \label{nJP}
\Psi^{nJP}_{\delta \gamma}(\vec{k \ }) = \sum_{LSm_Lm_S} \langle L m_L S m_S
\arrowvert J m_J \rangle
(-1)^{\frac{1}{2}+\gamma} \langle \frac{1}{2} \delta \frac{1}{2} -\gamma
\arrowvert S m_S \rangle
Y^{m_L}_L(\hat{k}) \Psi^{nJP}_{LS}(k)
\end{equation}
where again the color index is omitted.
Note the phase factor and negative magnetic substate sign in the Clebsch-Gordan
coefficient due to the transformation properties of
antiparticles under the $SU(2)$ rotation group. A thorough discussion
is given in Ref. \cite{Eisenberg}.

Inserting Eq. (\ref{nJP}) into Eq. (\ref{TDA}) yields the TDA
partial-wave equation of  motion
\begin{eqnarray*}
&(E_{nJP}-E_0-2\epsilon_k)\Psi^{nJP}_{LS}(k)(2J+1) =
\sum_{\Lambda \Sigma m_\Lambda m_\Sigma m_J  m_L  m_S}
\langle Jm_J \arrowvert L
m_L Sm_S \rangle \langle \Lambda m_\Lambda \Sigma m_\Sigma \arrowvert Jm_J
\rangle \\
&\int d\Omega_kd\Omega_q Y^{*m_L}_L(\hat{k})
Y^{m_\Lambda}_\Lambda(\hat{q}) \frac{4}{3}\int_0^\infty \frac{q^2dq}{(2\pi
)^3}
\Psi_{\Lambda \Sigma }^{nJP}(q)\hat{V}(\arrowvert \vec{k}-\vec{q}\arrowvert )\\
&
\sum_{\alpha \beta \gamma \delta} (-1)^{1+\beta + \gamma}
h^{\alpha \beta}_{\gamma \delta}(k,q) \langle S m_S
\arrowvert
\frac{1}{2} \alpha \frac{1}{2} -\beta \rangle \langle \frac{1}{2} \delta
\frac{1}{2}-\gamma \arrowvert \Sigma m_\Sigma \rangle
\end{eqnarray*}
where the function $h^{\alpha \beta}_{\gamma \delta}(k,q)$ contains the
gap angle from contractions involving rotated
spinors
\begin{eqnarray*}
&h^{\alpha \beta}_{\gamma \delta}(k,q)= \frac{1}{4} [ c_kc_q
(\delta_{\alpha \delta}g_{\gamma \beta} +\delta_{\gamma \beta}g_{\alpha
\delta}) +  &(1+s_k)(1+s_q)\delta_{\alpha
\delta}\delta_{\beta \gamma} +(1-s_q)(1-s_k)g_{\alpha \delta}g_{\gamma \beta}]
 \end{eqnarray*}
with
$$g_{\alpha \beta} = \chi^\dagger_\alpha \,\vec{\sigma}\cdot
\hat{q} \,
\vec{\sigma}\cdot
\hat{k} \, \chi_\beta \ . $$
Denoting the meson mass for state $nJP$ by $M_{nJP} = E_{nJP} - E_0$
and using the multipole expansion formulas for the interaction
yields the final TDA equation appropriate for numerical calculation
\begin{equation}
\label{Tammgeneral}
(M_{nJP}-2\epsilon_k)\Psi^{nJP}_{LS} (k)=\sum_{\Lambda \Sigma}\int_0^\infty
K^{JP}_{L \Lambda S \Sigma}(k,q) \Psi^{nJP}_{\Lambda \Sigma}(q)
\frac{q^2dq}{12\pi^2} \ .
\end{equation}
Note the Hamiltonian spin dependence generates a kernel that couples different
orbital and spin states.

We now apply these equations to the low lying meson spectrum with quantum
states specified by $I^G(J^{PC})$ having  $C$ parity,  $C =
(-1)^{L+S}$, and $G$ parity, $G = (-1)^{L+S+I}$.  In our model we neglect
the small electromagnetic (isospin violating) effects as well as coupling to
the gluon sector so that $I = 0$ and $1$ states are degenerate for the same
$J^{PC}$. For pseudoscalar states, $J^{PC} = 0^{-+}$,
$S=L=J=0$ giving  only one wavefunction component (no coupling).
This is also the case for scalar mesons ($L=1, S=1$)
having $J^{PC} = 0^{++}$. However
for the vector meson sector   $J^{PC} = 1^{--}$  both $(L=0,
S=1)$ and $(L=2, S=1)$ waves are allowed and, in general, will be coupled. 
Similarly for low lying pseudovector mesons having $(L=1, S=0$ or $S=1)$ and tensor
mesons,
with $(L=1, S=2)$ and $(L=3, S=2)$, there will be coupled equations.  Although
these equations are not difficult to solve, we only include the lowest orbital
partial-wave component and neglect all coupling since it has been computed
small
for the harmonic oscillator potential \cite {Lisbon}. There is then only
one kernel
$K(k,q)$ for each meson state with structure:

\begin{itemize}
\item  pseudoscalar, $L=S=J=0$, $$K(k,q)=2( c_k c_q \hat{V}_1
+(1+s_ks_q)\hat{V}_0)$$
\item scalar, $L=S=1$, $J=0$, $$K(k,q)=2(c_kc_q \hat{V}_0
+(1+s_ks_q)\hat{V}_1)$$
\item vector,  $L=0$, $S=J=1$ (neglecting the tensor $L=2, S=J=1$
coupling),
 $$K(k,q)=2c_kc_q\hat{V}_1 +(1+s_k)(1+s_q)\hat{V}_0
+(1-s_q)(1-s_k)(\frac{4\hat{V}_2-\hat{V}_0}{3})$$
\item pseudovector, $L=J=1$, $S=0$ (degenerate with $L=S=J=1$),
$$K(k,q)=c_kc_q(\hat{V}_0+\hat{V}_2)+2(1+s_ks_q)\hat{V}_1$$
\item tensor, $L=S=1$, $J=2$ (neglecting $L=3, S=J=2$ coupling),
$$K(k,q)=c_kc_q(3\hat{V}_2-\hat{V}_0)+(1+s_k)(1+s_q)\hat{V}_1 +
(1-s_k)(1-s_q) \frac{12\hat{V}_3-7\hat{V}_1}{5} \ . $$ 
 \end{itemize}
Note for the pseudovector mesons the kernels for $S =0$
and $S = 1$ are identical which differs from
Ref.
\cite{Orsay}.

We have also applied our approach to other flavored ($s$ and $c$) meson
systems and have obtained similar, but more complicated TDA
equations.  As a representative result, consider the pseudoscalar
$D$ meson with a $u$ (or
$d$) and $c$ quark.  The gap equation for the $c$ quark remains the same,
except for current mass, now $1.2 \ GeV$, which gives a different gap
energy, $\epsilon^c_k$, and angle, $s^c_k$. The TDA
equation, however, has a different form and generalizes to

\begin{eqnarray*}
\lefteqn{ (M_D- \epsilon^u_{k} - \epsilon^c_{k}) \Psi_{D}(k)
=  \frac{1}{3}\int\frac{q^2dq}{(2\pi)^2}\Psi_{D}(q) \cdot} \\ &\cdot
\left[ \left( \sqrt{1+s^u_{k}}\sqrt{1+s^u_{q}}\sqrt{1+s^c_{k}}
\sqrt{1+s^c_{q}} +\sqrt{1-s^u_{q}}\sqrt{1-s^u_{k}}
\sqrt{1-s^c_{k}}\sqrt{1-s^c_{q}} \right ) \hat{V}_0(k,q) \right. \\
&\left. \left( \sqrt{1-s^u_{q}}\sqrt{1-s^u_{k}}\sqrt{1+s^c_{k}}
\sqrt{1+s^c_{q}} + \sqrt{1+s^u_{k}} \sqrt{1+s^u_{q}}
\sqrt{1-s^c_{k}}\sqrt{1-s^c_{q}} \; \right)\hat{V}_1(k,q)\right] \
\end{eqnarray*}
with obvious form for other mixed flavors.
All equations are finite for $k = q$ as the IR divergence terms
from the confining potential again cancel.
We have also derived and solved the TDA equations for other spin
parity states which is further detailed in Sec. V.

\subsection{RPA and the quasiboson approximation}

The TDA can be improved by
utilizing a better vacuum with additional quasiparticle correlations
beyond the BCS.  Consistent with many-body applications in other disciplines
we now formulate the RPA \cite {Ring,Mattuck} and introduce a new vacuum,
$\arrowvert \Omega _{RPA} \rangle$, having both fermion
(two quasiparticles or Cooper pairs) and boson (four quasiparticles
or meson pairs) correlations.
The RPA meson state
\begin{equation}\label{RPAansatz}
\arrowvert \Psi^{nJP}_{RPA} \rangle =Q_{nJP}^{\dagger}(RPA) \arrowvert
\Omega_{RPA} \rangle
\end{equation}
involves a meson creation operator which is a generalization of
Eq. (\ref{TDAop})

\begin{equation}\label{RPA1}
Q_{nJP}^{\dagger}(RPA)=\sum_{ \lambda \mu}
\int \frac{d\vec{k}}{(2\pi)^3} [X ^{nJP}_{\lambda \mu}
B^{\dagger}_{\lambda}(\vec{k}\ ) D^{\dagger}_{\mu} (-\vec{k}\ )
- Y^{nJP}_{\lambda \mu}
B_{\lambda}(\vec{k}\ ) D_{\mu} (-\vec{k}\ )] \ .
\end{equation}
The RPA vacuum  then satisfies
$$Q_{nJP}(RPA) \arrowvert
\Omega_{RPA} \rangle  = 0$$
which, because of additional correlations from admixtures
of particle-hole excitation states, is not true for the
BCS vacuum.

To derive the RPA equations of motion we use Eq. (\ref{TDA})
and replace the BCS vacuum with $\arrowvert
\Omega_{RPA} \rangle$ and also substitute $\Psi^{nJP}_{RPA}$ for
$\Psi^{nJP}_{TDA}$ to generate one equation for the $X$ component.
We then repeat, changing the commutator to $[H,B_\alpha D_\beta ]$
to obtain the $Y$ equation.
Following standard treatments in other fields of
physics, we also invoke the quasiboson approximation and treat the fermion pair
operator $BD$ as a pure boson operator.  This significantly reduces the
commutator algebra complexity and generates one of the two coupled
equations for
the RPA wavefunctions $X$ and $Y$.

For the important pseudoscalar meson channel we obtain for the excited
state $n$

\begin{eqnarray}\label{RPAeq}
2\epsilon_k X^{n}(k) + \frac{1}{3} \int_0 ^{\infty} \frac{q^2dq}
{(2\pi)^2} \left[ X^{n}(q)F(k,q) + Y^{n}(q) G(k,q)
\right] = M_{n} X^{n}(k)
\\
2\epsilon_k Y^{n}(k) + \frac{1}{3} \int_0 ^{\infty} \frac{q^2dq}
{(2\pi)^2} \left[ Y^{n}(q)F(k,q) + X^{n}(q) G(k,q)
\right] = -M_{n} Y^{n}(k)
\end{eqnarray}
where

\begin{eqnarray}
F(k,q)=2c_qc_k \hat{V}_1 + 2(1+s_qs_k)\hat{V}_0\\
G(k,q)=2c_qc_k \hat{V}_1 - 2(1-s_qs_k)\hat{V}_0 \ .
\end{eqnarray}
Similar expressions directly follow for the other spin-parity states.

We adopt the standard normalization for the RPA wavefunctions
\begin{equation}
\langle \nu ' \arrowvert \nu \rangle = \langle \Omega_{RPA} \arrowvert
Q_{\nu'} Q_{\nu}^\dagger \arrowvert \Omega_{RPA} \rangle = \delta_{\nu\nu'}
\end{equation}
yielding
$$ \int_0^\infty k^2dk (X^{\nu '}(k)^*X^\nu(k)-Y^{\nu '}(k)^*Y^\nu(k))=
(2\pi)^3\delta_{\nu \nu'} \ . $$

The RPA equations, which
reduce to the TDA equations
in the limit $Y$ or $G$ $\rightarrow 0$,
are again an eigenvalue problem for $M_{nJP}$ which
can be easily diagonalized.
Related, the matrix size
can be reduced by a factor of 2 using the variables $X+Y$
and $X-Y$.
Finally, the equations are also IR finite for the sigular point $k = q$.

\subsection{Weak decay constants}

A crucial test of any approach is the ability to describe hadronic
decays.  In this paper we compute weak decays and defer our analysis
of hadronic decays to a subsequent publication.  
For a pseudoscalar meson $P$ with momentum $p_{\mu}$, energy $E_P$, mass $M_P$, the
weak decay constant,
$f_{P}$,  is defined for our normalization by
\begin{equation} \label{fdecay}
\langle \Omega \  \arrowvert  A_{\mu}(0)  \arrowvert
P(\vec{p} \ ) \rangle =   \frac{1}{\sqrt{E_P }} { f_P p_{\mu}}\ .
\end{equation}
Here $A_{\mu}(\vec{x}) = \overline{\Psi}(\vec{x}) \gamma_{\mu}\gamma_5
\Psi(\vec{x})$ is the axial current which specifies the chiral charge operator
\begin{equation}
Q_5 = \int d\vec{x}A_{0}(\vec{x}) =\int d\vec{x} \Psi^\dagger (\vec{x}) \gamma_5 
\Psi (\vec{x}) \ .
\end{equation}
Simplifying Eq. (\ref{fdecay}) for a meson at rest  yields
\begin{equation} \label{decay}  f_P=\frac{1}{\sqrt{M_P }} \langle
\Omega \arrowvert \Psi^{\dagger}(0)\gamma_5\Psi(0) \arrowvert P(0) \rangle \ .
\end{equation}
Applying this result for the TDA pion wavefunction gives the TDA pion
decay constant
\begin{equation} \label{f1}
f^{TDA}_{\pi}= \frac{1}{\pi \sqrt{(2\pi)^3M_{\pi}}}\int_0^{\infty}
\Psi^{\pi}_{TDA} (q) s_q \  q^2dq \ .
\end{equation}
Similarly, for the RPA pion wavefunction we obtain
\begin{equation} \label{pdecay}
f^{RPA}_{\pi}=\frac{1}{\pi \sqrt{(2\pi)^3M_{\pi}}}\int_0^\infty
 s_q(X^\pi(q)+Y^\pi(q))  \ q^2dq \ .
\end{equation}

Our results easily generalize to the $SU_f(3)$ flavor nonet. Now there
are nine axial charges given by
$$Q_5^a=\int d\vec{x}A^a_{0}(\vec{x}) =\int d\vec{x} \Psi^\dagger (\vec{x}) \gamma_5
\frac{\lambda^a}{2} \Psi (\vec{x})$$ where the eight Gell-Mann $\lambda^a$ matrices
are supplemented by
$\lambda^0=\sqrt{\frac{2}{3}}I$ to obtain both the octet and the singlet
under $SU_f(3)$ transformations.  The appropriate generalizations of Eq. (\ref{f1})
are then:
$$f^{TDA}_K=\frac{1}{2\pi\sqrt{(2\pi)^3M_K}}\int_0^\infty
{q^2dq}\Psi^K_{TDA}(q)
\cdot ( \sqrt{1+s^s_{q}} \sqrt{1+s^u_{q}} -\sqrt{1-s^s_{q}}\sqrt{1-s^u_{q}})$$
$$f^{TDA}_{\eta_8}=\frac{1}{3\pi\sqrt{(2\pi)^3M_{\eta_8}}}\int_0^\infty
\Psi_{TDA}^{\eta_8} (s^u_{q}+2s^s_{q}) {q^2dq}$$
$$f^{TDA}_{\eta_0}=\frac{1}{3\pi\sqrt{(2\pi)^3M_{\eta_0}}}\int_0^\infty
\Psi _{TDA}^{\eta_0} (2s^u_{q}+s^s_{q}) {q^2dq} \ .$$
We will use the above results in the next subsection to derive a
generalized Gell-Mann-Oakes-Renner relation and also in Sec. V
were we report numerical results.

\subsection{Chiral symmetry and the Gell-Mann-Oakes-Renner relation}

Chiral symmetry is a significant element of hadronic QCD and should be
present in all realistic models. Even though our vacuum properly exhibits
dynamic chiral symmetry breaking, our model Hamiltonian does indeed
respect this symmetry since  the commutator
$$[H, Q_5]  = m \int d\vec{x} \Psi^\dagger (\vec{x})  \Psi
(\vec{x}) \simeq 0 $$
essentially vanishes, consistent with the small $u$, $d$ quark mass.
Related, our RPA states also preserve chiral symmetry as the
RPA meson creation operator commutes with the chiral charge
in the chiral limit ($m \rightarrow 0$)
$$ [Q^{\dagger}(RPA), Q_5] = 0 \ .$$
However, the TDA operator,
Eq. (\ref{TDAop}) above, does not commute with $Q_5$
and violates chiral symmetry
since it is not fully
symmetric in operator structure (only contains
$B^{\dagger}D^{\dagger}$ and not $BD$).
This can also be documented by chiral transforming the TDA meson state
verifing that $B^\dagger D^\dagger$ rotates to combinations of $B^\dagger B$,
$DD^\dagger$ and $DB$. Hence, the TDA ansatz is not closed under a chiral
rotation and
Goldstone bosons will not appear in the TDA spectrum.  We therefore expect
significant, but unphysical, chiral symmetry violations in the TDA calculations and
anticipate the TDA pion mass to be much larger than in the RPA which is confirmed
in Sec. V as only the RPA calculations yield a Goldstone pion
in the chiral limit.

On the other hand, the RPA ansatz, Eq. (\ref{RPA1}), is chirally invariant
since it is
closed under this  rotation in the chiral limit ($\arrowvert X
\arrowvert =
\arrowvert Y \arrowvert$).
Hence the  mechanism of spontaneous symmetry breaking is present and the pion
mass  is zero according to its Goldstone boson nature
as we numerically verify in the next section.

With these results we now derive two different Gell-Mann-Oakes-Renner
(GMOR) relations,  one based upon exact model eigenstates while
the other relates RPA states and the BCS vacuum.  Assume we have the
complete set of exact eigenstates, $\arrowvert
n \rangle$, to our QCD model Hamiltonian, including the vacuum ground state
$\arrowvert
\Omega_{exact} \rangle$.  Evaluating the double commutator

\begin{equation}
\langle \Omega_{exact} \arrowvert \left[ Q_5,\left[Q_5,H\right] \right]
\arrowvert
\Omega_{exact} \rangle =
4m \langle  \overline{q}q\rangle_{exact}
\end{equation}
then generates the exact quark condensate.  Evaluating the
double commutator again, but now invoking twice the
completeness relation, $1=\sum_n \arrowvert n \rangle \langle n
\arrowvert $, and identifying the decay constant relation,  Eq. (\ref {decay}),
leads to the generalized GMOR relation

\begin{equation} \label{GMOR}
-2m\langle \overline{q}q\rangle= \sum_{n} M_n^2 f_n^2
\end{equation}
summed over all (ground and excited) pseudoscalar meson states with mass $M_n$
and decay constants $f_n$.

This can be extended to flavor using two of the nine $SU_f(3)$ axial charge
operators,
$Q^a_5$ and $Q^b_5$, to derive the following GMOR relations:
\begin{itemize}  \item a= b = 1, 2, or 3
$$-2 m_u\langle \overline{u} u \rangle = \sum_{\pi^n} M^2_{\pi^n}
f^2_{\pi^n}$$ \item a = b = 4, 5, 6, or 7
$$-\left( \frac{m_u+m_s}{2} \right) \langle \overline{u} u + \overline{s} s
\rangle = \ \sum_{K^n} M^2_{K^n} f^2_{K^n}$$
\item a = b = 8,
$$-\left( \frac{2}{3} m_u \langle \overline{u} u \rangle + \frac{4}{3}
m_s \langle \overline{s} s \rangle \right) = \sum_{\eta^n_8}
f^2_{\eta^n_8}M^2_{\eta^n_8}$$
\item a = b = 0
$$-\left( \frac{4}{3}m_u\langle \overline{u} u \rangle +\frac{2}{3} m_s
\langle \overline{s} s \rangle \right) = \sum_{\eta^n_0}
f^2_{\eta^n_0}M^2_{\eta^n_0} \ . $$
\end{itemize}

Because the exact eigenstates will generally not be available, these
equations are of limited value. However, they do provide testing
criterion for approximation solutions.  Further, since decay constants
are suppressed for excited states $n$ we can drop higher terms to obtain
more useful relations involving ratios such as
$$ \left(\frac{M_Kf_K}{M_\pi f_\pi} \right)^2 = \frac{m_u+m_s}{2m_u}
\left( \frac{\langle \overline{u}u + \overline{s}s \rangle}{2\langle
\overline{u}u \rangle} \right) $$
which we will use in Sec. V in connection with our discussion
of the kaon mass calculation.

Next we utilize Thouless' theorem \cite{Thouless} applied to the
chiral charge operator
\begin{equation}
\langle \Omega \arrowvert \left[ Q_5,\left[Q_5,H\right] \right]
\arrowvert
\Omega \rangle =
2 \sum_{n}\arrowvert\langle\Psi^{nJP}_{RPA} \arrowvert Q_5 \arrowvert
\Omega_{RPA} \rangle
\arrowvert^2(E_n-E_0)_{RPA}
\end{equation}
to immediately derive the RPA GMOR relation
\begin{equation}
-2m\langle \overline{q}q\rangle_{BCS}=  (M_{\pi}^2 f_{\pi}^2 )_{RPA}
\end{equation}
where we have again dropped the excited state decay constants.
Note that the left hand side entails the BCS vacuum while the right hand side
involves the RPA states and energies.  This relation clearly predicts
the RPA pion is a Golstone boson in the chiral limit
(i.e $M_{\pi} \rightarrow 0$
for $m \rightarrow 0$) which we numerically confirm in the next section.

Finally,  calculating the RPA meson mass spectrum does
not require obtaining the RPA vacuum (see Eq.(\ref{RPAeq})),  but computing the
RPA decay constants does. Determining $\arrowvert \Omega_{RPA} \rangle$ is
actually quite difficult, however, the leading correction can be
approximately calculated
using another theorem by Thouless \cite{Ring}
\begin{equation} \label{RPAvacuum}
\arrowvert \Omega_{RPA}\rangle \simeq \arrowvert \Omega \rangle +
(F^\dagger)^2
\arrowvert \Omega \rangle \end{equation}
where $F^{\dagger}$ is a TDA type operator given by

$$F^\dagger = \sum_{\alpha\beta}\int {d\vec{q}}
f_{\alpha \beta}(\vec{q})^*
B^{\dagger}_{\alpha} (\vec{q})D^\dagger_{ \beta}(-\vec{q})  $$
with the fermion pair operator
$B^\dagger D^\dagger$ now obeying  bosonic commutation relations.
Here $f^*$ is an unknown amplitude assumed to be
small (this is not true in the chiral limit as discussed in the next section).
The approximate RPA vacuum is thus described as a mixture of the BCS
vacuum having Cooper pairs (Eq. \ref{BCS}) and
two quasibosons (mesons) coupled to vacuum quantum numbers ($0^{++}$).
Since, as
shown in the next section, the RPA is of importance for mainly the low-lying
pseudoscalar states, we assume that only these states  contribute to
$\arrowvert
\Omega_{RPA}
\rangle$ and neglect all others. Our
result easily generalizes to scalar or other mesons (appropriately coupled to
$J^P = 0^{++}$). Within this approximation we obtain an improved
quark condensate to be  compared with Eq. (\ref{cond})

$$ \langle \overline{q}q \rangle_{RPA}\simeq\langle \overline{q}q
\rangle_{BCS} +
\frac{8\sum_{\alpha\beta}\int {d\vec{q}}
 s_{q}
\arrowvert f_{\alpha \beta}(\vec{q})\arrowvert^2}{1+2\sum_{\alpha\beta}\int
{d\vec{q}}
\arrowvert f_{\alpha \beta}(\vec{q})\arrowvert^2} \ . $$ 

To obtain $f_{\alpha \beta }$, we impose
$Q_\pi (RPA) \arrowvert \Omega_{RPA} \rangle =0$ and use Eq.
(\ref{RPAvacuum})   neglecting higher order terms corresponding to Fock
states with more than two pions. We also
only retain ground state meson (pion)  contributions
from $X$ and $Y$ yielding

\begin{equation} \label{solvef} f_{\alpha \beta}(q)=
N^{-1} Y(q)(-1)^{{1\over 2}+\alpha+\beta} \langle \frac{1}{2} -\alpha
\frac{1}{2} \beta \arrowvert 00\rangle
\end{equation}
with a normalization constant depending on both RPA wavefunction components
$$N^2=2\int \frac{d\vec{q}}{(2\pi)^3} Y(q) X(q) \ . $$
The improved RPA quark condensate is

\begin{equation}
\langle \overline{q}q \rangle_{RPA} \simeq \langle \overline{q}q \rangle_{BCS}
+ c \int_0^\infty
s_qY(q)^2  q^2 dq
\end{equation}
where the constant c is given by
$$c = \frac{2\int Y(q)^2 q^2dq }{(\int Y(q)X(q)q^2dq)^2 +\frac{1}{2} (\int
Y(q)^2q^2dq)^2} \ . $$
 Also the improved pion decay constant to this
vacuum is
\begin{equation}
f_\pi\simeq \frac{1}{\pi\sqrt{(2\pi )^3 M_\pi}}\int s_q(X(q) +
c\pi^2 N^2 Y(q))q^2dq
\end{equation}
in contrast to Eq. (\ref{pdecay}). We have found that for meson masses
above 800 $MeV$ $Y$ is very  small and there is no essential difference 
between RPA and TDA.

We now comprehensively apply the above formulas and conduct a comparative
analysis of the TDA and
RPA approaches.

\section{Applications and numerical results}

In this section we first present and discuss our TDA spectra  for the
pseudoscalar and vector mesons and then in subsection B we compare
to our RPA results for both meson masses and  the decay constants.
Subsection C treats mixing for the $\eta$ and $\eta'$ mesons while
subsection D details applications to charmed mesons, especially
$J/\Psi$ states.

\subsection{Tamm-Dancoff spectrum}

Solving the TDA eigenvalue problem is straight forward. The diagonal part of
the matrix contains the
IR singularity that is rigorously controlled by cancellation,
again permitting numerical regulation  by simply
skipping the point
$q=k$ as in the gap equation solution. Due to the linear nature of this
system, results are convergent for a mesh as sparce as 700 points.

Tables I and II
summarize the resulting TDA meson spectrum corresponding to the five $JLS$
kernels specified in Sec. IV A.
The energy
difference  between  pseudoscalar and vector states is about 200 $MeV$ for
the $u$/$d$
quark mesons,  80 $MeV$ for the open flavored
($K$, $K^*$) and only 50 $MeV$ for
the pure strange composites ($\phi$). Our one
parameter model can not accurately
describe the entire observed splitting indicating
additional dynamics beyond simple
spin interactions from a Dirac spinor field is needed. In
general, the masses are in good agreement with the PDG \cite {PDG00} accepted values
for the vector mesons, but the TDA  low-lying
pseudoscalar sector is deficient. 
This is expected since in this channel  vacuum (chiral
symmetry)
effects  are most prominent. In the scalar channel the situation is more
confusing, since other hadron states, some with explicit gluonic structure, can
more easily mix. Since our unified
model allows us to treat  glueballs, mesons and hybrids comprehensively,
future  work will further address  understanding this channel. Interestingly,
our lightest $f_0$ mass, which has a P wave oribital excitation, is below
1 $GeV$. We also find that the computed TDA masses are not
very sensitive to
details of the vacuum that enters via
the gap function characterizing the BCS ground state.

As described above, we may also use a simplified TDA equation to extract a
constituent quark mass. In the chiral
limit, the mass obtained is 51 $MeV$ and this dressing is roughly constant,
consistent with $SU_f(3)$ symmetry,
up to current masses of 150 $MeV$, where the generated constituent mass
is 203 $MeV$. We are therefore dealing with light dressed quarks, even for
strange quarks which may prove a deficiency when calculating
electromagnetic form factors. We shall also address this issue in the near
future.

Another important feature of our relativistic effective
Hamiltonian  is that it naturally includes the
kinematical and spin dependent interactions (e.g. spin-spin,
spin-orbit,
tensor).  These effects are
very important in the light quark sector even after chiral symmetry
breaking, because the light quark constituent mass generated in our
scheme, around 80 $MeV$, is still small as compared to our interaction
scale (424 $MeV$).
In
particular, notice in Table II the spin splitting between the $0^{++}$,
$1^{++}$
and
$2^{++}$ mesons, all having the same $L$ and $S$ quantum numbers (also observe
the large radial excitation in each  channel). The level spacing is consistent
with  Ref.
\cite{Orsay} but very different from the  naive expectations from the
constituent quark model. The is due to the confining
(nonperturbative) $\vec{L}\cdot \vec{S}$ coupling,  which is only of order
$\alpha_s$ in the
constituent quark model. The spin spacing is governed by
the matrix element
 $$\langle LSJ \arrowvert \vec{L}\cdot \vec{S}
\arrowvert LSJ \rangle = \frac{1}{2} [J(J+1) - L(L+1) - S(S+1)] $$
which, for $L = S = 1$, reduces to $-2+\frac{J(J+1)}{2}$.
The $-2$ in this expression describes the light scalar meson,
whereas the $J(J+1)$ contribution explains the splittings in Table II.
The calculated
$2^{++}$ mass is much  heavier than the lightest observed $f_2$ at 1270 
$MeV$. Clearly, including coupling to the $L=3$, $S=1$ channel as well as
multi-quark Fock states will alter, but improve, our prediction.
Further, and as mentioned previously, additional quark-gluon interaction terms
should also be included in our Hamiltonian which have a different Lorentz
spin structure.  In our model this would generate a weaker hyperfine
interaction
of order $\alpha_s$. This would also provide additional splitting
between
$S = 0$ and $S = 1$ levels that would improve our TDA, and especially RPA,
$\pi$-$\rho$ mass underprediction.  Such an analysis would fully
clarify the relative importance of chiral symmetry versus spin interactions
as the latter is generally attributed the dominant effect in conventional
constituent quark models having color magnetic, effective one
gluon exchange potentials (see Ref. \cite{deRujula}).
We are currently examining this  issue and will report
results in a future paper.

Finally and also related  is the  spin-orbit
splitting  for other flavors which is summarized in Tables I and IV.
In Table IV we illustrate the familiar $J(J + 1)$ dependence by fitting the
TDA spectrum for different flavors.

\subsection{RPA spectrum and decay constants}
We now present our RPA results and compare with both TDA and observation.
Table III and Figs. 2 through 7 highlight our key results.  In general the RPA
and TDA masses agree except for the light pseudoscalar mesons where the
RPA provides a better description.  This is because, as
discussed above, only the RPA correctly implements chiral symmetry
as illustrated in Fig. 3 where the lightest scalar and
pseudoscalar masses are plotted as a continuous function of the current quark
mass.  Note that only the pseudoscalar mass approaches zero in the
chiral limit consistent with Goldstone's theorem (the observed pion
mass is reproduced for a $u$ quark mass of about 2 $MeV$).

The kaon system reveals the largest model deficiency 
(see Table I). Even the RPA in the chiral limit produces a
too massive kaon, about 850 $MeV$.  As indicated by Fig. 4, to
reproduce the observed kaon requires a strange quark mass of
about 50 $MeV$.
A detailed
analysis reveals that the explicit contribution of the current quark mass
to the RPA equation -through its appearance in $\epsilon(k)$- is only
additive. It is the gap angle, introducing an implicit flavor
dependence, which inhibits a lighter kaon mass. We could
fit both pion and kaon masses by adjusting the current quark masses, but we
prefer  awaiting improvements from renormalizing the quark gap
equation (work in progress and will be subsequently publish).  The gap angle for
a non-zero current quark mass is very sensitive at  high momentum to
the dominanting Coulomb potential  and sizeable corrections are expected.
Also for higher lying  excited meson states there is also the issue of two particle,
two hole Fock state contributions.

In Fig. 5 we compare our scalar, pseudovector and tensor meson TDA and
RPA predictions to data.  In general there is qualitative agreement.
Note that for the higher excited states above 1 $GeV$ the
TDA and RPA results are identical and it is clear that these
systems are not
governed by chiral symmetry.

We also illustrate the behavior of various wavefunctions.
In Fig. 6 we detail the difference between TDA wavefunctions
for the $f_0$ and $\pi$ mesons. Figure 7 compares the TDA and RPA
wavefunctions for the pion.

Finally we discuss decay constants.  Both the
TDA and RPA pion decay constants are too small, about 17 $MeV$, in contrast
to the observed value of 93 $MeV$.
This is
consistent with the Orsay \cite{Orsay} results.  We again attribute
this to the
model Hamiltonian and vacuum as the BCS angle does not have sufficiently
high momentum components.
However, if we use the approximate RPA vacuum in
the quasiboson approximation (see Eq. (4.25)) the improved RPA
decay constant increases to 57 $MeV$.  Appropriately, the condensate also
significantly increases to $-(320 \, MeV)^3$, in much better agreement
with lattice results $\approx -(250 \,MeV)^3$.  The latter result is
consistent with applications in nuclear
physics where the RPA tends to overcorrelate the ground state.

Since we have only approximately evaluated the RPA vacuum
by truncating at the two pion level it is not surprising that $f_\pi$
does not agree with measurement (or chiral perturbation theory
results) and therefore needs further refining. We expect the truncation
to the two pion level, Eq. (\ref{RPAvacuum}), to be reasonable provided we
are not in the chiral limit, since then $Y \ll X$. However, we would like to be
able to calculate in that regime. Further, the calculation in Ref.
\cite{Lisbon2} points to a necessary decrease of the BCS condensate
when including coupled channels. They  argue that for
a pure chiral pion, coupling additional channels would decrease its mass
which might even become negative, destabilizing the vacuum. This
argument seems
sound and since we are above the chiral quark mass limit, at a
model value $m=5$ $MeV$ where our pion is too massive (277 $MeV$), this
decrease is a welcome improvement.
We defer further discussion until publication by our collaborative effort with
the  Lisbon group which will also clarify  pionic correlations
in the  ground state of our approach.

Recalling that  our computed condensates require renormalization except in
the chiral limit, we can only test the GMOR relation for $m = 0$ which
is trivially satisfied in the RPA. However, we note that
the decay
constants for excited pion states are much smaller and rapidly approach zero in
the chiral limit. Hence the GMOR relation is satisfied with
predominatly the first state. 
This is not true for heavier quarks and to numerically satisfy the GMOR requires
including
several eigenstates.

\subsection{$\eta-\eta'$ mixing}

An extremely challenging but still not understood problem is
the $\eta$, $\eta'$ system and attending flavor mixing of the light quarks.
Although our effective Hamiltonian has an explicit flavor dependence through the
current quark masses, it still conserves flavor.  However, if the gluon sector
is included, such as through the hyperfine, minimal coupling interaction, an
effective flavor dependence naturally emerges though higher order 
quark-glue-quark effects and dynamic mixing of flavor states is possible.  We
are currently deriving such a term which is similar, but more rigorous than
the t'Hooft interaction based upon instantons (classical glue).  This will
be reported in a future communication, however, it is still of interest
to perform a simple $\eta$, $\eta'$  mixing analysis by introducing a
flavor off-diagonal interaction as we now detail.

With no dynamic flavor mixing, the $\eta$ and $\eta'$ are (poorly) modeled
as  $SU_f(3)$ octet, $\eta_8$, and singlet, $\eta_0$, states
respectively given by (we adopt the convention used in Refs. \cite{gilman,kks}) 
\begin{eqnarray}
\eta&\approx\eta_8=&\sqrt{\frac {1} {3}}\,\, n\overline{n} - \sqrt{\frac {2}
{3}} \,\,  s\overline{s} = \cos \theta_{ SU_f(3)} \,\, n\overline{n}  - \sin
\theta_{SU_f(3)}
\,\,s\overline{s} \\
\eta'&\approx\eta_0=&\sqrt{\frac {2} {3}}\,\,n\overline{n} + \sqrt{\frac {1}
{3}} \,\,  s\overline{s}  = \sin
\theta_{SU_f(3)} \,\, n\overline{n}  + \cos
\theta_{SU_f(3)}
\,\,s\overline{s}
\end{eqnarray}
involving the isoscalar
$n\overline{n} = (u\overline{u} +d\overline{d})/\sqrt{2}$ and $s\overline{s}$
states and pure $SU_f(3)$ mixing angle $\theta_{SU_f(3)} = 54.74^{\circ}$.  We
have computed the TDA masses of the pure $n\overline{n}$ and
$s\overline{s}$ meson states to be $M_n = $ 612 $MeV$ and $M_s =$ 1002 $MeV$,
respectively.  Hence the predicted, pure $SU_f(3)$, $\eta$, $\eta'$
masses are
\begin{eqnarray}
M_{\eta}&\approx M_{\eta_8}=& \cos^2 \theta_{SU_f(3)} \,\, M_n  + \sin^2
\theta_{SU_f(3)}
\,\,M_s = 872 MeV\\
M_{\eta'}&\approx M_{\eta_0}=& \sin^2 \theta_{SU_f(3)} \,\, M_n  + \cos^2
\theta_{SU_f(3)}
\,\,M_s = 742 MeV .  
\end{eqnarray}
For the RPA, $M_n = $ 290 $MeV$ and $M_s =$ 978 $MeV$, yielding
$M_{\eta} = 749$ $MeV$ and $M_{\eta'} = 519$  $MeV$.   
Similarly the $SU_f(3)$ $\eta$, $\eta'$ decay constants are given
in terms of the $u/d$ isoscalar, $f_n$, and strange quark, $f_s$, decay
constants
\begin{eqnarray}
f_{\eta}&\approx& f_{\eta_8}=\cos^2 \theta_{SU_f(3)} \,\,f_n  + \sin^2
\theta_{SU_f(3)}
\,\,f_s \\
f_{\eta'}&\approx&f_{\eta_0}=\sin^2 \theta_{SU_f(3)} \,\, f_n  + \cos^2
\theta_{SU_f(3)}
\,\,f_s .
\end{eqnarray}
Note the quadratic dependence on angles due to expressing both the axial current and
meson states in their $SU_f(3)$ representations. 
Using the TDA computed values of $f_n = 17$ $MeV$ and $f_s = 75$ $MeV$, yields
$f_{\eta_8} = 56$ $MeV$ and $f_{\eta_0} = $ 40 $MeV$. The RPA values are
$f_n = 57$ $MeV$ and $f_s = 75$ $MeV$, giving
$f_{\eta_8} = 68$ $MeV$ and $f_{\eta_0} = $ 63 $MeV$.

We can now improve these results by generalizing our Hamiltonian, which is
diagonal in flavor space, to have off-diagonal matrix elements. The simplest
prescription is to  just add 
a constant, $\langle q\overline{q} 
\arrowvert H' \arrowvert q \overline{q} \rangle = \lambda$, to both
diagonal, $\langle n\overline{n} 
\arrowvert H' \arrowvert n \overline{n} \rangle = 2\lambda$,
$\langle s\overline{s} 
\arrowvert H' \arrowvert s \overline{s} \rangle = \lambda$,  and off-diagonal
$\langle n\overline{n} 
\arrowvert H' \arrowvert s \overline{s} \rangle = \sqrt{2}\lambda$,
terms giving

\begin{equation}
H = \left( \begin{array}{cc}
  M_n + 2\lambda & \sqrt{2}\lambda \\ \sqrt{2}\lambda & M_s +
\lambda
 \end{array} \right ) .
\end{equation}
Diagonalizing $H$ leads to the new, mixed mass eigenvalues 
\begin{eqnarray}
M_{\eta}&=&\frac{M_n+M_s+3\lambda}{2}-\frac{1}{2}
\sqrt{M_n^2+M_s^2+9\lambda^2+2(\lambda M_n -M_sM_n -M_s \lambda)} 
\\ M_{\eta'}&=& \frac{M_n+M_s+3\lambda}{2}+\frac{1}{2}
\sqrt{M_n^2+M_s^2+9\lambda^2+2(\lambda M_n -M_sM_n -M_s \lambda)}
\end{eqnarray}
and eigenstates
\begin{eqnarray}
\eta&=& \cos (\theta_{SU_f(3)}+ \theta_P) \,\, n\overline{n}  - \sin
(\theta_{SU_f(3)}+ \theta_P)
\,\,s\overline{s}=\cos\theta_P \,\,\eta_8 - \sin \theta_P \ \,\,\eta_0
\\
\eta'&= &\sin (\theta_{SU_f(3)}+ \theta_P) \,\, n\overline{n}  + \cos
(\theta_{SU_f(3)}+ \theta_P) \, \, s\overline{s} =\cos\theta_P \,\,\eta_0 + \sin
\theta_P \ \,\, \eta_8
\end{eqnarray}
involving rotation by an additional angle $\theta_P$ that is a function of
$\lambda$.   The mixed, presumably more physical, decay constants are then
\begin{eqnarray}
f_{\eta}&=& \cos \theta_P \, \, f_{\eta_8} - \sin\theta_P \, \, f_{\eta_0}\\
f_{\eta'}&=&\cos \theta_P \, \, f_{\eta_0} + \sin\theta_P \, \, f_{\eta_8}.
\end{eqnarray}
Performing a least squares fit to the observed masses ($M_{\eta} =$ 547 $MeV$,
$M_{\eta'} =$ 958 $MeV$), yields
$\lambda = - 33 \, MeV$ ($\theta_P =$ $-61^{\circ}$) for the TDA, which in
turn produces $M_{\eta} =$ 541 $MeV$, $M_{\eta'} =$ 974 $MeV$,  $f_{\eta}  =$ 62
$MeV$ and $f_{\eta'}=$ -30 $MeV$.  For the RPA, $\lambda =$ 82 $MeV$ ($\theta_P
=$
$-44^{\circ}$) generating $M_{\eta} =$ 433 $MeV$, $M_{\eta'} =$ 1081 $MeV$, 
$f_{\eta}  =$ 93 $MeV$ and $f_{\eta'}=$ -3 $MeV$.  It is interesting that  
while the simple mixing provides improvement, the TDA masses are in better agreement
than the RPA. We could also improve the decay constants utilizing a two angle
mixing formalism \cite{feldemann} but refrain since clearly 
a more sophisticated treatment is necessary which will be provided by our
quark-gluon coupling formulation in the near future.

\subsection{Heavy mesons}
The constituent quark models and non-relativistic expansions of QCD offer
more reliable  results for  heavy quark systems
where physical intuition from Quantum Mechanical bound states is more appropriate.
Hence the charmed mesons afford a good limiting testing for our 
relativistic approach.
We again calculate the spectrum of the charmed mesons (see Figures
[\ref{UCspectrum},\ref{SCspectrum}]) and of charmonium (Fig.
[\ref{CCspectrum}]) using our many-body model. The TDA
is now sufficient since
chiral symmetry, crucial for the light mesons, is not a constraint
and the RPA will produce the same results.
Using a charmed quark mass of 1200 $MeV$, the general features of the
spectra are well reproduced and the radial excitations, $\Psi (2S)$ and
$\eta_c$, are adequately described. We therefore expect our predictions
for the remaining unconfirmed states to be reasonable.

The angular momentum splittings of these systems are known to be
dominated by the one gluon exchange potential (OGE) which we have not included
in this calculation. Hence there will be improvement from future calculations
based upon our
renormalized project.
A general feature reflected by our charmed spectra is the near vanishing of the
spin-spin interaction, leading to degenerate $0^{-+}$ and $1^{--}$ states.
Since the 100-150 $MeV$ hyperfine splittings in these systems
will presumably be recovered when we include the perturbative OGE, we
do not comment further.
As for the spin-orbit splittings, our results are too large for the $D$
mesons and too small for the $\chi_c$ mesons. These splittings are
adequately explained in non-relativistic quark models (see
Ref. \cite{N. Isgur}) where the absence of a large spin-orbit 
effect for light quark masses is attributed to the cancellation between the
Thomas precession in the confining potential and the one gluon exchange effective
potential, although it is a bit concerning that the actual
splitting between the $\chi_{c0}$ and the $\chi_{c1}$ is twice the size
of the
$\chi_{c1}$-$\chi_{c2}$ splitting, when according to the spin-orbit
$J(J+1)$ rule, it should be a half.

Finally, we note that by including D waves in the charmonium spectrum
we are able to resolve the long standing "overpopulation" problem of
$J/\Psi$ states relative to observation.  This is clearly illustrated
in Fig. 10.  This (previous) deficiency in number of $c\overline{c}$ states
has been characterized as evidence for glueballs and/or hybrid
mesons because the $J/\Psi$
system is believed to be gluon rich.  Our result suggests, however,
that simple
level counting may not be effective in identifying hadrons with explicit
gluonic degrees of freedom.

\section{Outlook and summary}
Before summarizing our results we comment on the strengths and weaknesses
of our many-body approach as well as some attending, open hadronic physics issues.
Beginning with our Hamiltonian, $H$, the current-current (density-density) color
interaction  forbids free, isolated colored objects in the theory. For
the vacuum this is realized in the BCS
by an infinite shift in the free quark self-energy 
due to the integral of
$\hat{V}(\arrowvert
\vec{k}-\vec{q}
\arrowvert)$. Similarly for hadrons in both TDA and RPA, 
colored composite objects (e.g. diquarks) are precluded by
the appearance of $\hat{V}(0)$ which is divergent, whereas 
in the singlet channel this divergence is removed
by vanishing color factors.
Next we note that $H$ conserves chiral symmetry yet our BCS vacuum properly
exhibits dynamic chiral symmetry breaking. Further, our RPA pion emerges as a
Goldstone boson in the chiral limit. This is not true for the TDA pion since only
the RPA excitation operator commutes with the chiral charge. Another,
significant model feature is that  Fock state truncation is a controlable
approximation amendable to systematic improvement.
Thus our
Hamiltonian many-body approach is an attractive, promising method
for comprehensively investigating hadronic structure 
as it embodies 
confinement, chiral symmetry breaking and orderly
construction of multiparticle Fock states. It also provides an
excellent vehicle for testing more fundamental effective Hamiltonians as well
as affording a powerful phenomenological framework for hadron structure.
Considering the
form of $H$, with only a single predetermined dynamical parameter,
it is encouraging that the
chiral limit is adequately reached in the RPA and that the meson spectrum
is in qualitative agreement with experiment. To
achieve detailed, quantitative descriptions
will require further improvements in both the Hamiltonian and  effects
from including higher Fock
space components.  In particular, both the high energy behavior (Coulomb potential)
and quark-gluon coupling  effects (efectively instantons) will be incorporated
and reported in a future
publication. Finally,  our current study is similar,
but more extensive than the
Orsay analysis \cite{Orsay}
due to our application to multiflavor systems. Our results are also
more realistic (and numerically more difficult) then that work since
we have utilized a linear confining interaction, determined by lattice and
Regge phenomenology, which generates complicated nonlocal
integral equations, rather than solving a simpler differential equation for a
harmonic oscillator potential.

Summarizing, we have performed approximate, but large-scale diagonalizations of an
effective Coulomb gauge Hamiltonian utilizing standard many-body techniques. 
Using the BCS, a non-linear gap equation has been derived and accurately
solved to provide vacuum properties (quasiparticles and condensates).
Incorporating only  predetermined parameters (string tension, $\sigma$, and
reasonable current quark masses), we have qualitatively reproduced the
low energy $u,d,s$ and $c$ meson spectra. Most importantly,
we have obtained a chiral pion, detailed that chiral symmetry is responsible for the
large $\pi-\rho$ mass splitting and resolved the problem of overpopulation of
theoretical
$J/\Psi$ states.

Future work will address the full Hamiltonian in the combined quark and gluon
sectors. In particular, we will obtain improved
spin (hyperfine) and flavor (t'Hooft) interactions
from quark-glue coupling. This should provide  a better description of vacuum
properties and the scalar/pseudoscalar masses, especially the $\pi, K, \eta$ and
$\eta'$. We will also include more complex 2 quasiparticle-2 quasihole
Fock states 
for heavier $u/d$  mesons as well as 3 quasiparticles for  baryons and hybrids.
Much of this work is in progress and will  soon be reported.
 
\vspace{1cm}
\acknowledgments

The authors are grateful for comments and discussions with P. Bicudo, J. E.
Ribeiro, A. Szczepaniak and  members (present and former) of the NCSU theory group.
This work is supported in part by grants DOE DE-FG02-97ER41048 and NSF INT-9807009.
F. J. Llanes-Estrada was a SURA-Jefferson Laboratory graduate fellowship
recipient. Supercomputer time from NERSC is also acknowledged.

\newpage
\appendix
\section{BCS Vacuum State}
We further discuss the relation between the BCS rotated,
$\arrowvert \Omega \rangle$, and the trivial or perturbative, $\arrowvert 0
\rangle$, vacua.
We first note that the BCS vacuum state given by Eq. (\ref{BCS}) is
not a unitary transformation and
does not have a finite normalization.  This is because the operator
in the exponential is not antihermitian.   It
is therefore necessary to normalize  matrix elements by dividing with $\langle
\Omega
\arrowvert
\Omega
\rangle$ and this is implicit in our presentation. Alternatively, and
equivalently,
$\arrowvert
\Omega
\rangle$  can be represented by a norm preserving unitary transformation of the
form
$$
\arrowvert \Omega \rangle = e^{A^\dagger - A} \arrowvert \ 0 \rangle
$$
where
$$
A^\dagger = \sum_{ \lambda_1 \lambda_2} \int d\vec{k}\tan\theta_k
M_{\lambda_1 \lambda_2}b^\dagger_{\lambda_1}(\vec{k}\ )
d^\dagger_{\lambda_2}(-\vec{k}\ )
$$
and all flavor and color indices are suppressed.
Here $M_{\lambda_1 \lambda_2}$ are matrix elements of
the Pauli matrices
$M =  \vec{\sigma} \cdot
\hat{k}$
\begin{eqnarray} \nonumber
& & M_{\lambda_1 \lambda_2}
= \sum_\mu (-1)^{\mu} \hat{k}^{\mu}  {\sigma}^{-{\mu}}_{\lambda_1 \lambda_2}
\\ \nonumber
& &= \sqrt{8\pi} (-1)^{\frac{1}{2}-\lambda_2} \sum_{{\mu}\nu}   \langle \frac{1}{2}
\lambda_1 \frac{1}{2} -\lambda_2 \arrowvert 1 \nu \rangle 
 \langle 1 {\mu} 1 \nu \arrowvert 0 0 \rangle Y_1^{\mu}(\hat{k}) .
\end{eqnarray}

It is interesting to note that the BCS vacuum state is orthogonal to the
trivial vacuum, $\langle  0 \arrowvert \Omega \rangle = 0 $, in the infinite
volume limit.  Further, the Hilbert space vectors constructed from the two
different vacua are also orthogonal provided the BCS angle is nonzero
(vacuum condensates are present).  Because of this property the BCS rotation
has been called a pseudounitary transformation \cite{Lisbon}.  Consult
this reference for further details (note they have a different phase
convention and use $M =  \vec{\sigma} \cdot \hat{k} (i\sigma_2)$).

\newpage

\section{General TDA Equation}

Using the phase convention of Ref. \cite{Rose}, the general TDA meson
equation for arbitrary angular momentum is

\begin{eqnarray*}
(M_{nJP} -2 \epsilon_k) \Psi^{nJP}_{\Lambda\Sigma}(k)& =&
\int_0^\infty \frac{q^2dq}{\pi^2} \left[ \frac{(1+s_k)(1+s_q)}{12}
\hat{{V}}_\Lambda(k,q) \Psi^{nJP}_{\Lambda \Sigma}(q)  +\right.\\
 +\sum_{lLSf} \frac{c_kc_q}{2} \Psi^{nJP}_{LS}(q)
\hat{{V}}_l(k,q)
\prod_1 &+& \left.  \sum_{lLSfghL_1L_2} 3(1-s_q)(1-s_k)
\hat{{V}}_l(k,q) \Psi^{nJP}_{LS}(q) \prod_2
\right]
\end{eqnarray*}

where the angular momentum products are

\begin{eqnarray*}
\prod_1 = \langle 1 0 L 0 \arrowvert l 0 \rangle \langle 1 0 \Lambda 0
\arrowvert l 0 \rangle (2f+1) \sqrt{(2\Sigma +1)(2S+1)(2\Lambda
+1)(2L+1)} \\ \cdot  W(\Sigma \frac{1}{2} S \frac{1}{2};\frac{1}{2}f)
W(\frac{1}{2}1 \frac{1}{2}1;\frac{1}{2} f) W(L1\Lambda1;lf)W(LS\Lambda
\Sigma;Jf)(-1)^{J+L+1}(1+(-1)^{S+\Sigma})
\end{eqnarray*}
and

\begin{eqnarray*}
\prod_2 =
(2f+1)(2g+1)(2h+1)\sqrt{(2L_1+1)(2L_2+1)(2L+1)(2\Lambda+1)(2\Sigma+1)(2S+1)}
 \\ \cdot
 (-1)^{(1+2(f-h)+L-\Lambda+l-J)}\langle 1010\arrowvert L_10\rangle \langle
 1010\arrowvert L_20 \rangle \langle L_1 0\Lambda 0 \arrowvert 0\rangle
\langle L_20L0\arrowvert l0 \rangle  \\ \cdot
W(\frac{1}{2}\frac{1}{2}1L_1;1f)W(f\frac{1}{2}\frac{1}{2}\frac{1}{2};1\Sigma)
W(L_21\frac{1}{2}\frac{1}{2};fg)W(\frac{1}{2}\frac{1}{2}g\frac{1}{2};1S)
 \\ \cdot
W(L_2LL_1\Lambda;lh)W(SL\Sigma\Lambda;Jh)W(f\Sigma gS;\frac{1}{2}h)
W(L_2gL_1f;\frac{1}{2}h) .
\end{eqnarray*}
These formulas have been applied to several different
meson spin, parity states as presented in Sec. IV.

The moments of the angular integrations for the linear
potential are obtained from
$$\hat{V}_n^L=-8\pi\sigma\int_{-1}^1 \frac{1}{\arrowvert \vec{k} -\vec{q}
\arrowvert ^4} x^n dx$$
where $x=\hat{k}\cdot \hat{q}$. These can be calculated using the
recurrence relation

$$\hat{V}_n^L=\frac{k}{q} \hat{V}_{n-1}^L - \frac{4\pi\sigma}{q} \frac{d}{dk}
\left( -\frac{1}{2qk} \int_{-1}^{1}
\frac{x^{n-1}dx}{x-\frac{k^2+q^2}{2qk}} \right)$$
or by explicit integration

$$\hat{V}_n^L=\frac{k}{q} \hat{V}_{n-1}^L - \frac{4\pi\sigma}{2q^2}
\frac{d}{dk} \left[ \frac{1}{k} \left( \frac{x^{n-1}}{n-1}
+\frac{\omega x^{n-2}}{n-2} + \cdots + x \omega^{n-2} \right)_{-1}^1  +
\omega^{n-1} \log\left( \frac{\omega-1}{\omega+1}\right) \right]$$
where $\omega\equiv \frac{k^2+q^2}{2qk}$.
Evaluating the first three moments yields
$$\hat{V_0}^L=\frac{-16\pi \sigma}{(k^2-q^2)^2}$$
$$\hat{V_1}^L=\frac{2\pi\sigma}{k^2 q^2} \left[ ln \left( \frac{k+q}{k-q}
\right) ^2 +
(k^2+q^2)\left( \frac{-4qk}{(k^2-q^2)^2} \right) \right] $$
$$\hat{V_2}^L= 3\pi\sigma \frac{k^2+q^2}{k^3 q^3}
 ln \left( \frac{k+q}{k-q}
\right) ^2  - \frac{8\pi\sigma}{k^2 q^2} \frac{k^4+q^4}{(k^2 - q^2)^2} \ .$$

For the Coulomb potential, $V_C = \frac{-\alpha_s}{\arrowvert
\vec{x}-\vec{y}\arrowvert}$,
$\hat{V}_C=\frac{-4\pi \alpha_s}{\arrowvert \vec{k}-\vec{q}\arrowvert^{2}}$,
and the
angular integrals are

$$\hat{V_0}^C(k,q)=\frac{-2\pi\alpha_s}{qk} \ln\left(\left( \frac{k+q}{k-q}
\right)^2\right)$$
$$\hat{V_1}^C=\frac{4\pi\alpha_s}{qk}+\frac{k^2+q^2}{2qk}\hat{V}_0^C$$
$$\hat{V_2}^C=\frac{k^2+q^2}{2qk} \hat{V}_1^C .$$


\newpage

\begin{table} 
\caption{TDA ground and first excited states in $MeV$. Linear potential only,
$\sigma=0.18 \, GeV^2$.}
\label{TDAtable1}
  \begin{tabular}{|cc|cc|cc|cc|cc|}
    \multicolumn{1}{|c}{$$} & \multicolumn{1}{c|}{$$} &
    \multicolumn{2}{c|}{$0^{-+}$} & \multicolumn{2}{c|}{$0^{++}$} &
    \multicolumn{2}{c|}{$1^{- -}$} & \multicolumn{2}{c|}{$1^{+\pm}$}
\hspace{-.19cm}  
\\
\hline 
\multicolumn{1}{|c}{$m_1$} & \multicolumn{1}{c|}{$m_2$} 
     & $E_1$ & $E_2$ & $E_1$ & $E_2$ & $E_1$ & $E_2$ & $E_1$ & $E_2$ \\ \hline
    0 & 0 & 586 & 1473& 817& 1667& 798 &1602 & 1076& 1818  \\
    5 & 5 & 612 & 1494& 850& 1675& 800 &1615 & 1093& 1835 \\
    5 & 10 & 624&1503 & 861 &1703&  803& 1619& 1100& 1843\\
    5 & 150 & 877& 1679 &  1086 & 1873 &  957 & 1743 & 1273 & 1988\\
    150 & 150 & 1002 & 1808 & 1297 & 2044 & 1044 & 1849 & 1416& 2116 
  \end{tabular}
\end{table}

\begin{table} 
\caption{TDA $u/d$ mesons ($L=S=1$, no $L = 3$) in the chiral
limit.}
\label{TDAtable3}
   \begin{tabular}{|cccc|}
   $J^{\pi C}$ & $E_1$ & $E_2$ & $E_3$ \\ \hline
   $0^{++}$ & 817 & 1667 & 2301 \\   
   $1^{++}$ & 1076 & 1818 & 2411 \\
   $2^{++}$ & 1767  & 2281 & 2749 
   \end{tabular}
\end{table}

\begin{table} 
\caption{Chiral symmetry breaking in the RPA:
scalar vs. pseudoscalar spectrum.}
\label{RPAtable1}
  \begin{tabular}{|cc|cc|cc|}
    $m_1$ & $m_2$ & $E_1^{0++}$ & $E_2^{0++}$ & $E_1^{0-+}$ & $E_2^{0-+}$ \\
    \hline
    0 & 0 & 729 & 1652 & 0 & 1435 \\
    5 & 5 & 775 & 1679 & 300 & 1463 \\
    5 & 10 & 794 & 1641 & 350 & 1475 \\
    14 & 14 & 838 & 1719 & 441 & 1502 \\ 
    150 & 150 & 1288 & 2042 & 978 & 1805 
  \end{tabular}
\end{table}

\begin{table} 
\caption{TDA fits to the spin-orbit splitting, $E_{J+1}-E_J=AJ(J+1)$, for
$L=S=1$ as a function of the current quark mass.}
\label{TDAtable4}
   \begin{tabular}{|cccccc|}
    $m_1/m_2$ & 5/5 & 150/150 & 5/1200 & 150/1200 & 1200/1200 \\ \hline
    A ($MeV$) & 162 & 77 & 74 & 40 & 10 
   \end{tabular}
\end{table}

\begin{figure}
\psfig{figure=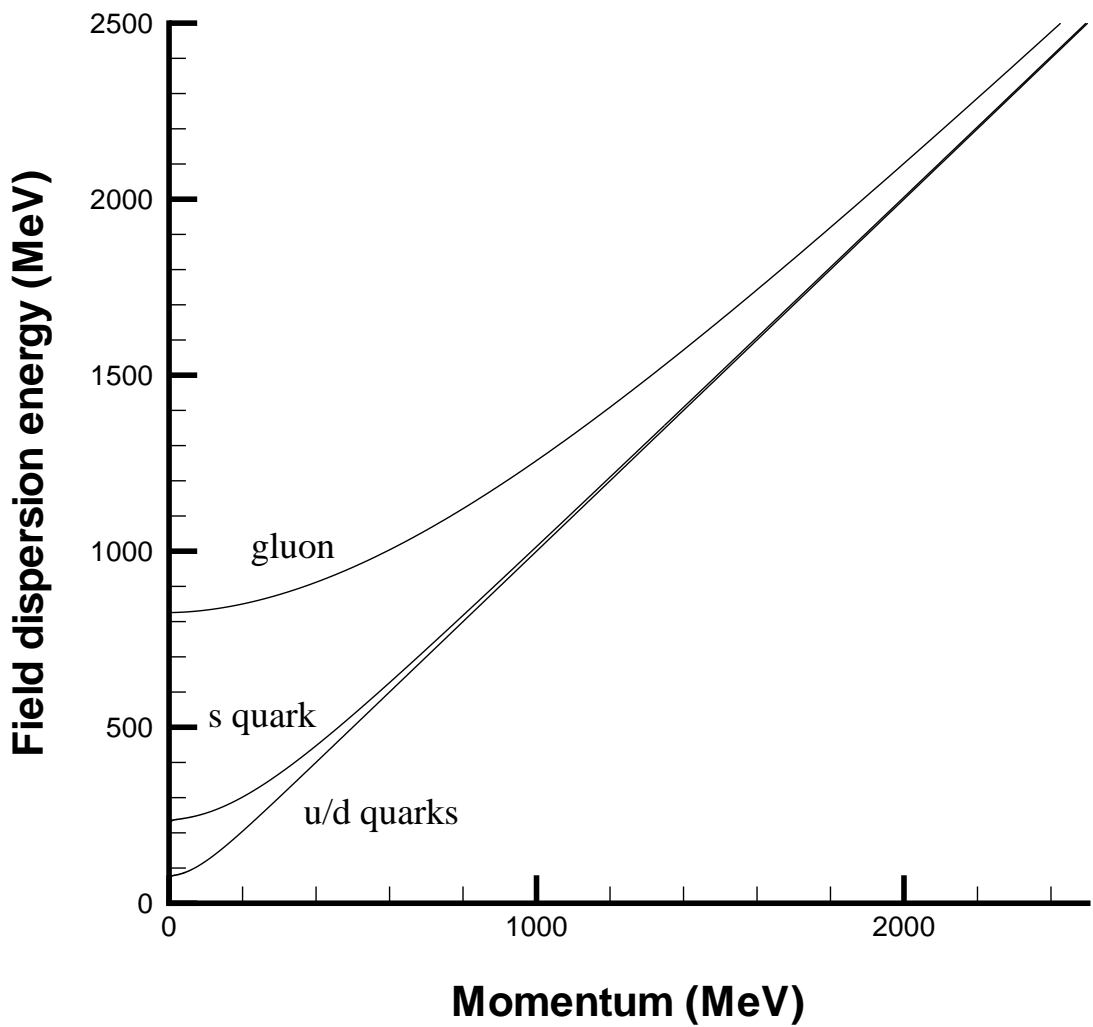,width=5in,height=7in}
\vspace{-.5in}
\caption{Quasiparticle energies for the $u/d, s$ quarks and
the gluon \protect\cite{ssjc96}.}
\label{dispersionfig}
\end{figure}

\begin{figure}
\psfig{figure=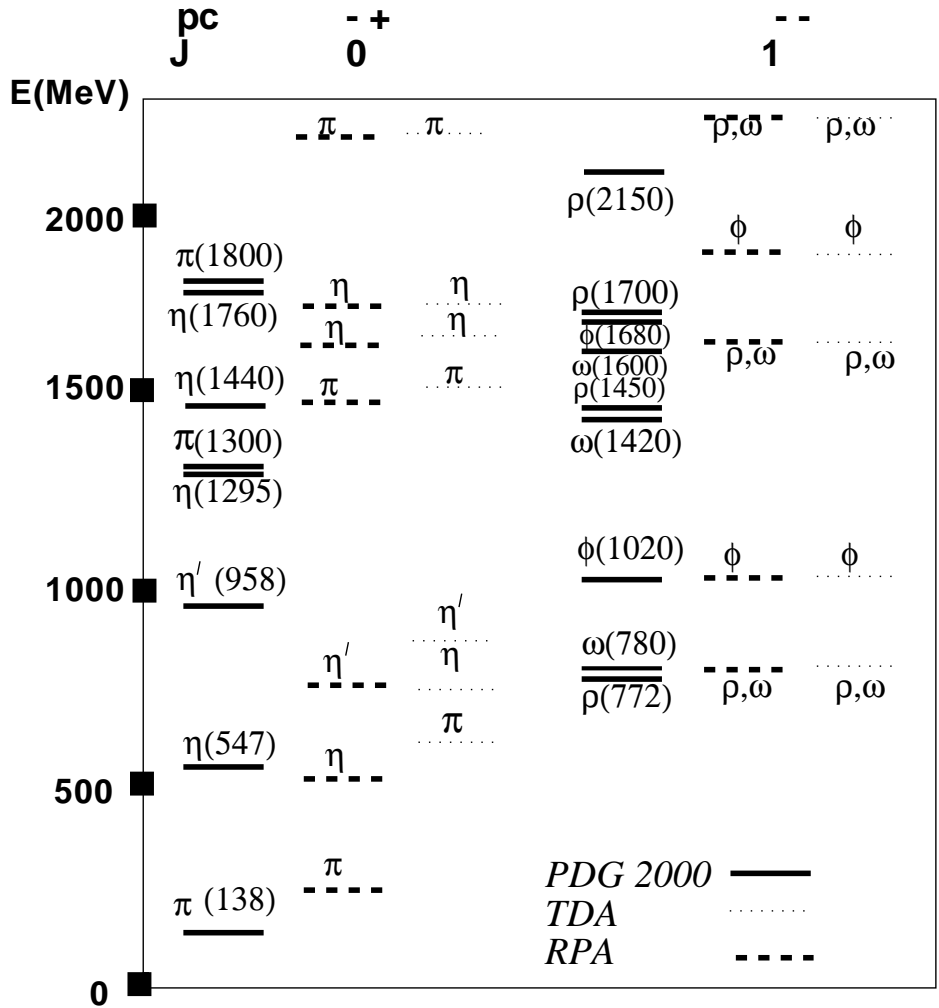,width=5.1in,height=7.25in}
\caption{Pseudoscalar and vector($L = 0$) TDA and RPA meson spectra. Only the
RPA provides the correct chiral limit for the pion.}
\label{L0spectrum}
\end{figure}

\begin{figure}
\psfig{figure=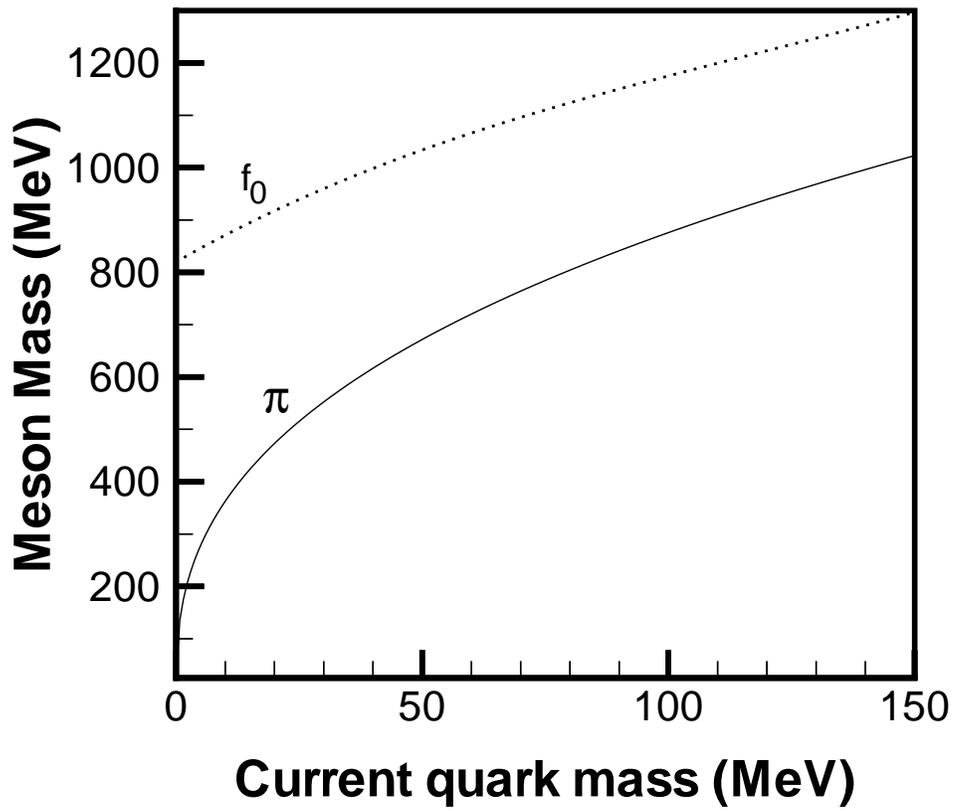,width=5.0in,height=6.8in}
\caption{Chiral symmetry in the
RPA. For $m \rightarrow 0$ the pseudoscalar
(solid) but not scalar (dotted) meson
mass vanishes.}
\label{CHIRALRPA}
\end{figure}

\begin{figure}
\psfig{figure=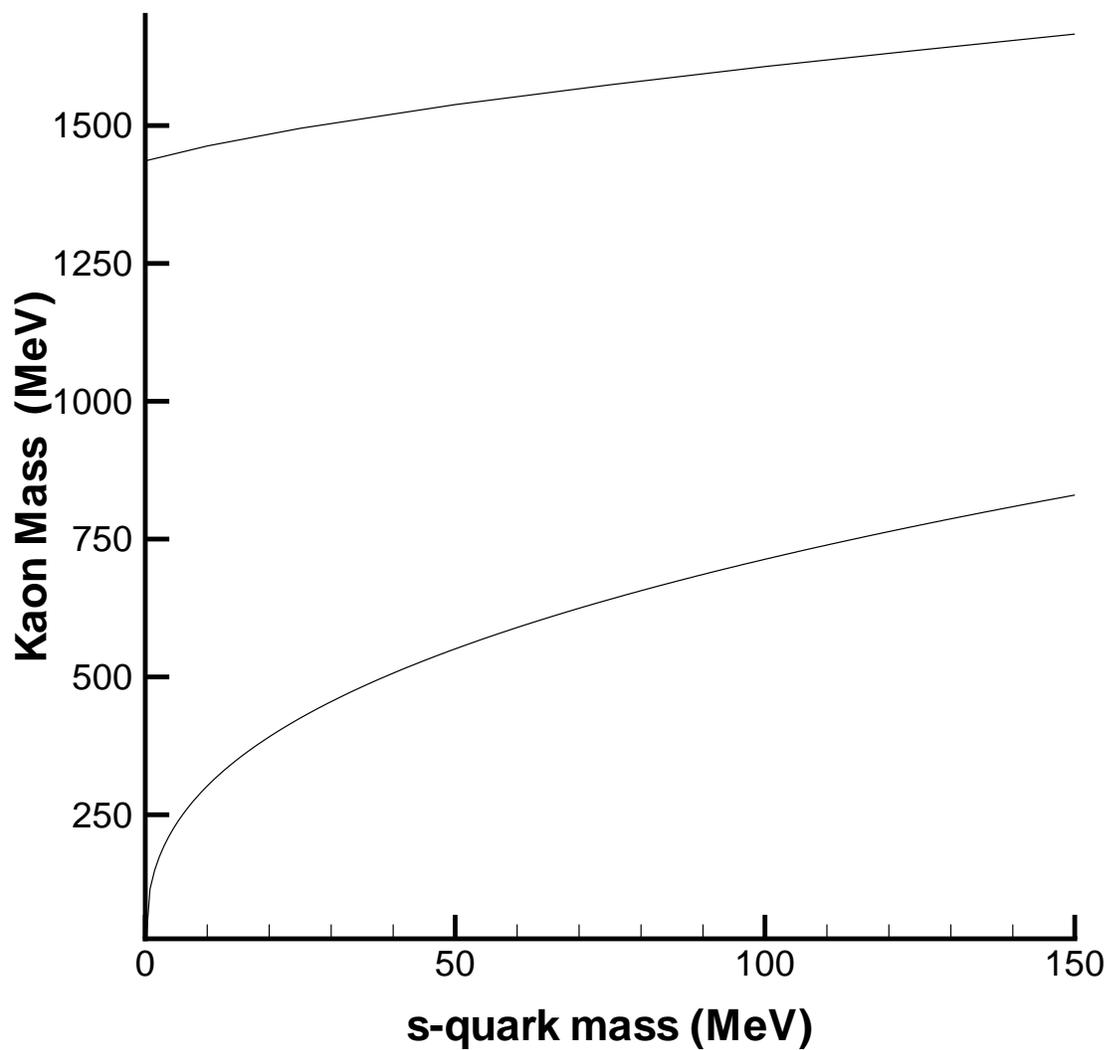,width=5in,height=5.5in}
\vspace{1.75in}
\caption{Chiral behavior of the kaon and first radial excitation. Light
quark mass is 0 $MeV$.}
\label{CHIRALKAON}
\end{figure}

\begin{figure}
\psfig{figure=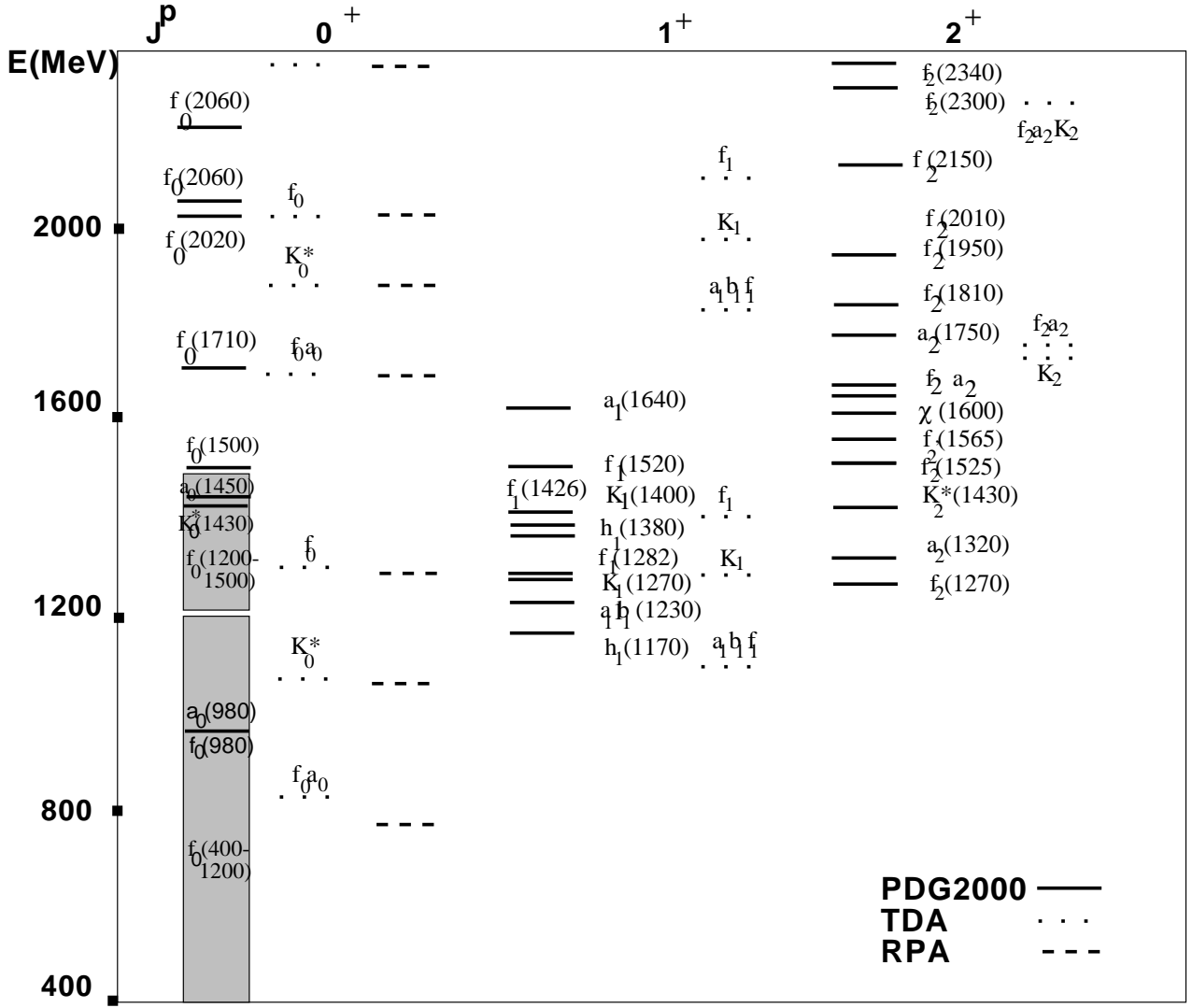,width=5.1in,height=7.2in}
\vspace{.5in}
\caption{Scalar,
pseudovector and tensor ($L = 1$) meson spectra. Above 1 $GeV$, the TDA (dots)
and RPA (dashes) are essentially identical.}
\label{L1spectrum}
\end{figure}

\begin{figure}
\psfig{figure=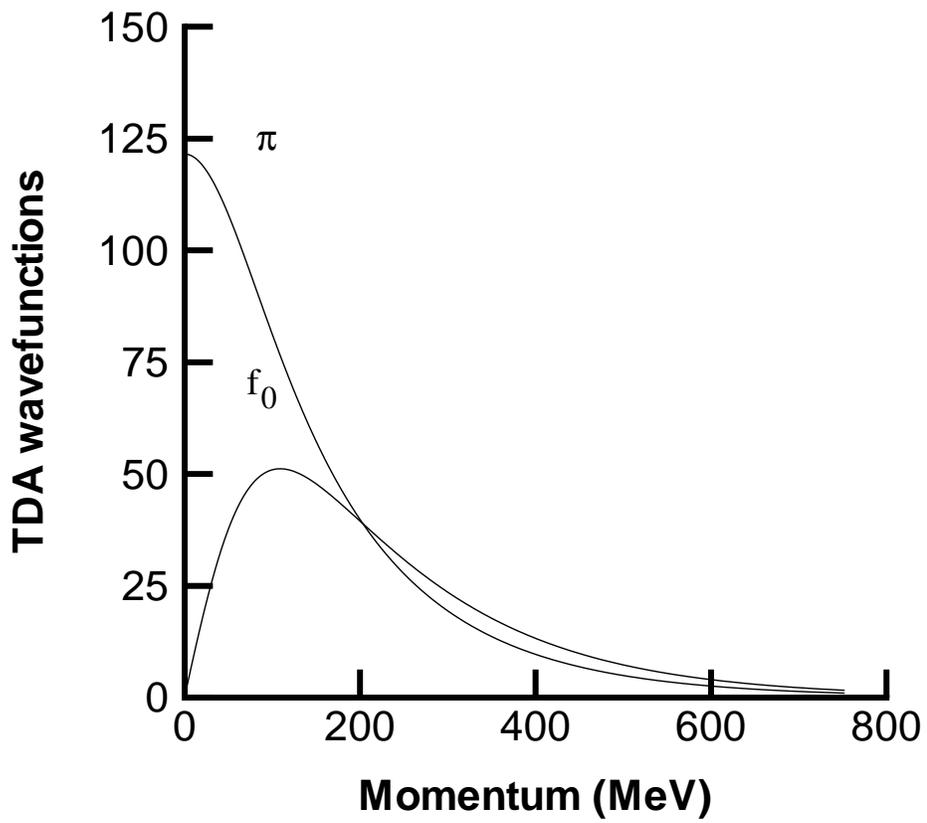,width=5in,height=8in}
\vspace{-.5in}
\caption{Ground state TDA wave functions for the pseudoscalar $\pi$ and scalar
$f_0$.}
\label{TDAwavefig}
\end{figure}

\begin{figure}
\psfig{figure=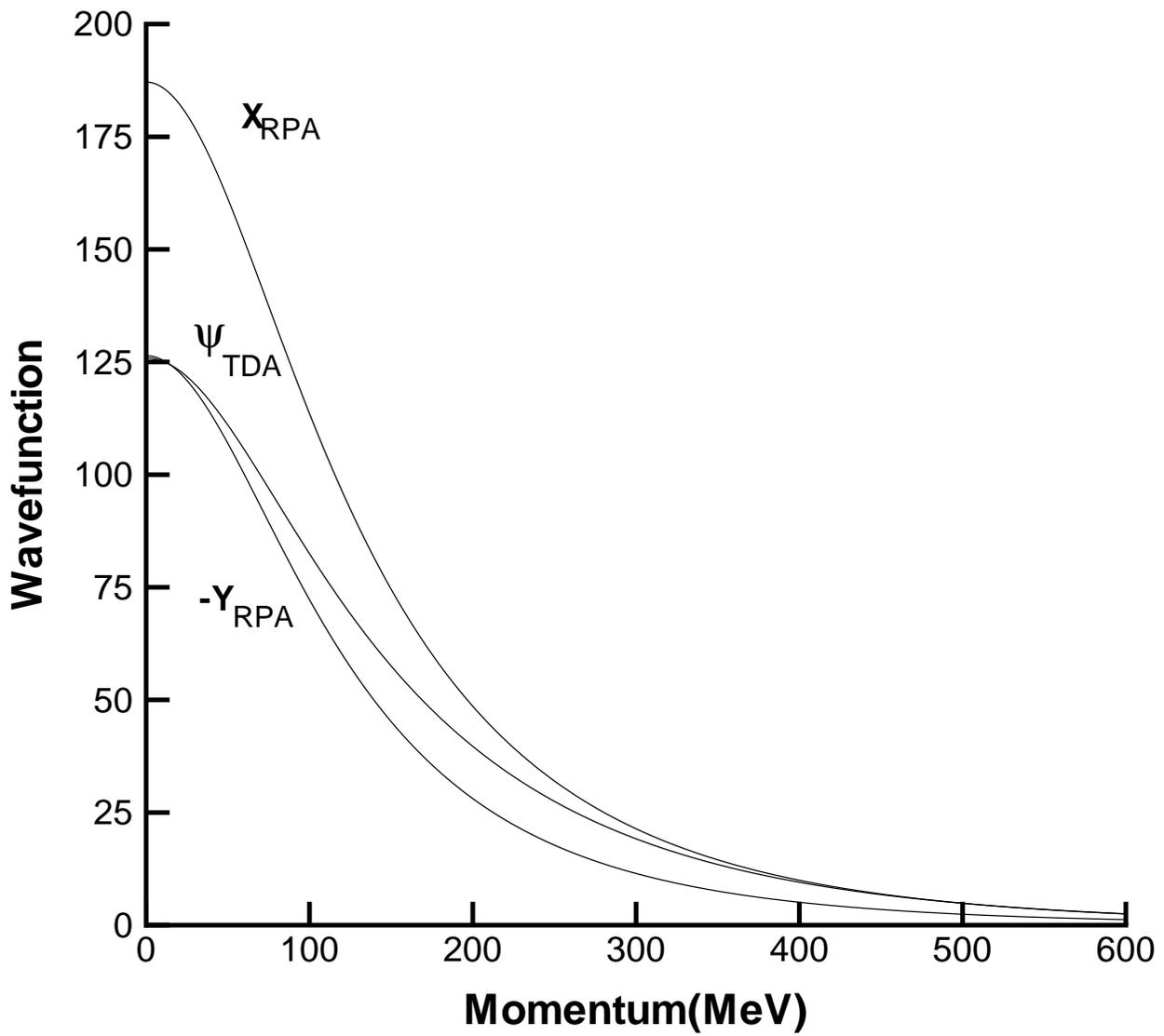,width=5in,height=7.5in}
\vspace {.75in}
\caption{TDA and RPA ground state pion wavefunctions.}
\label{TDARPAwavesfig}
\end{figure}

\begin{figure}
\psfig{figure=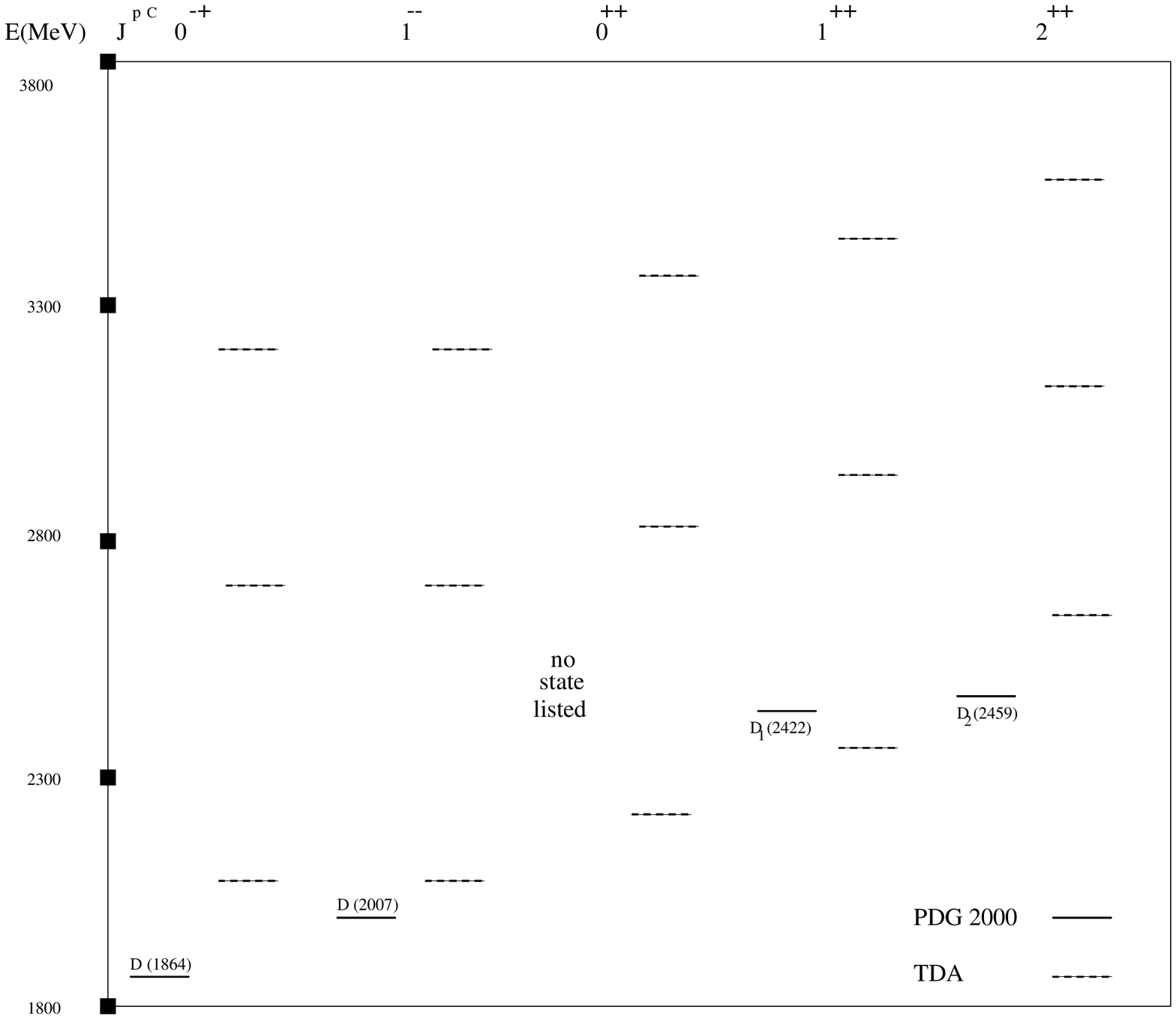,width=7in,height=7in}
\caption{$D$ mesons ($L=0$ and $L=1$). 
Charmed quark mass is 1200 $MeV$, light quark is 5 $MeV$.}
\label{UCspectrum}
\end{figure}

\begin{figure}
\psfig{figure=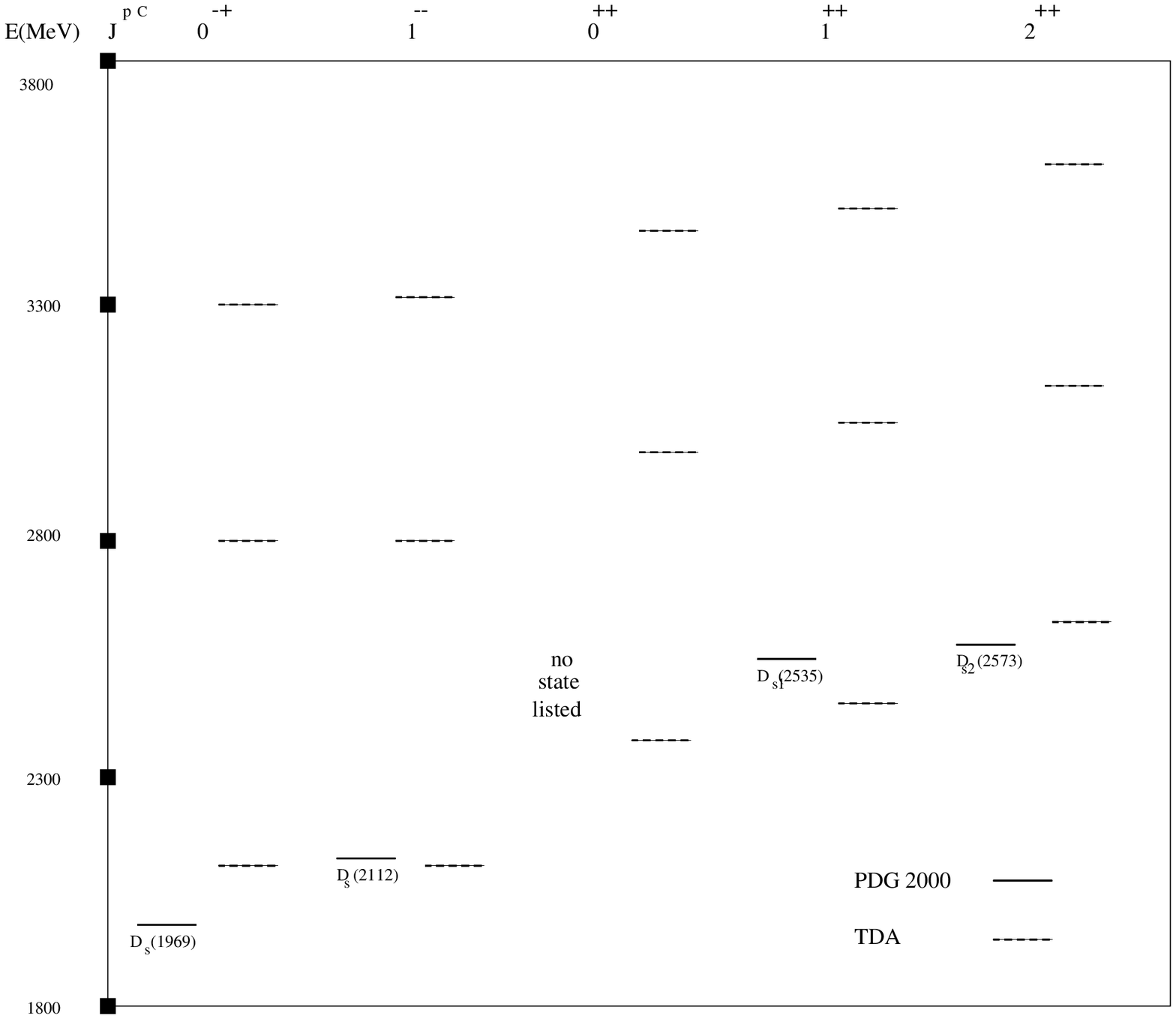,width=7in,height=7in}
\caption{$D_s$ mesons ($L=0$ and $L=1$).
Charmed (strange) quark mass
is 1200 (150) $MeV$.}
\label{SCspectrum}
\end{figure}

\begin{figure}
\psfig{figure=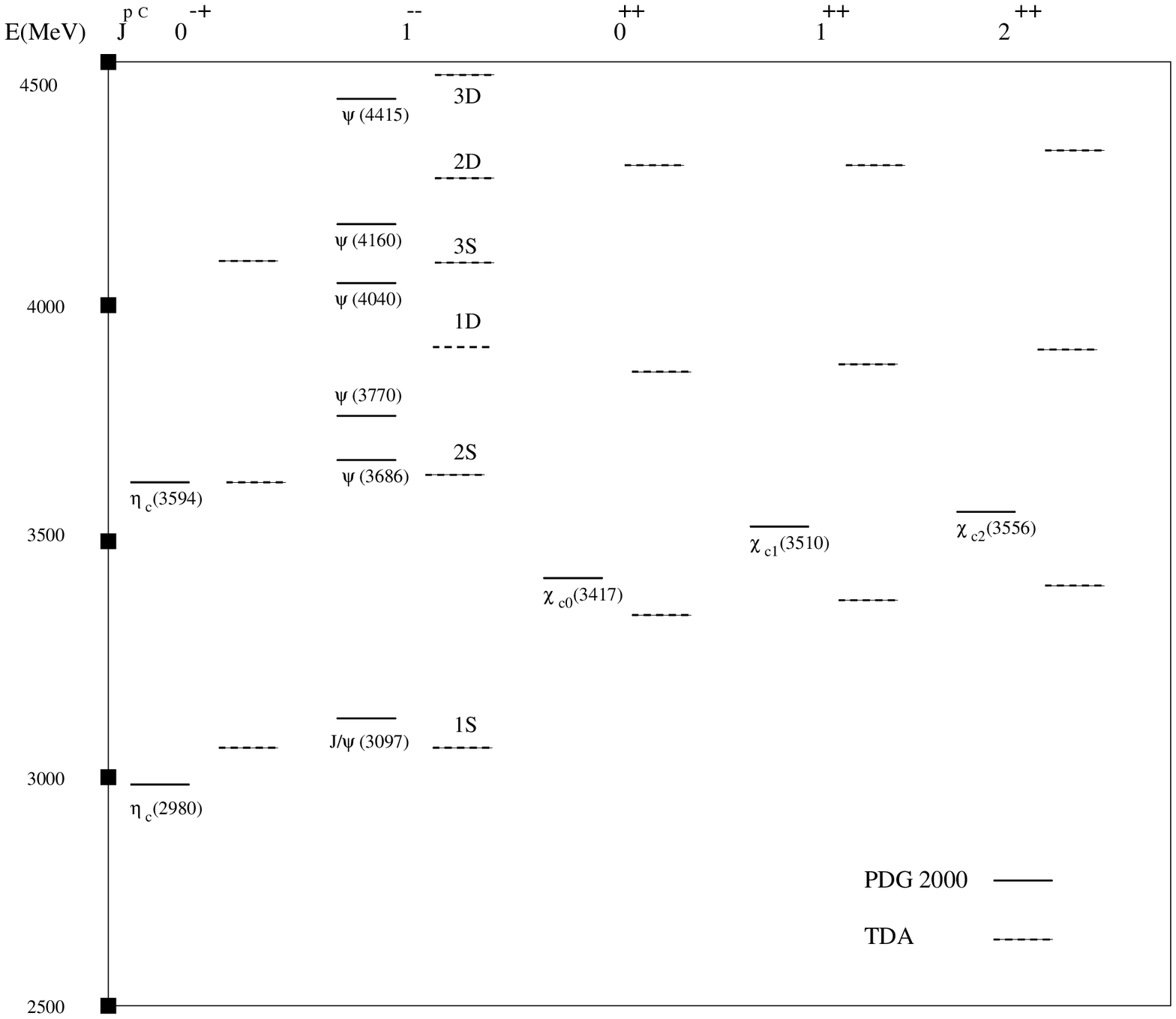,width=7in,height=7in}
\vspace {.3in}
\caption{Charmonium system ($L=0$ and $L=1$). 
Charmed quark mass is
1200 $MeV$.}
\label{CCspectrum}
\end{figure}

\end{document}